\documentclass[10pt,journal,compsoc]{IEEEtran}

\ifCLASSOPTIONcompsoc
  \usepackage[nocompress]{cite}
 \usepackage[caption=false,font=footnotesize,labelfont=sf,textfont=sf]{subfig}
\else
 \usepackage[caption=false,font=footnotesize]{subfig}
  \usepackage{cite}
\fi
\ifCLASSOPTIONcaptionsoff
 \usepackage[nomarkers]{endfloat}
\let\MYoriglatexcaption\caption
\renewcommand{\caption}[2][\relax]{\MYoriglatexcaption[#2]{#2}}
\fi
\usepackage{url}
\usepackage{fixltx2e}
\usepackage{enumitem}
\usepackage{array}
\usepackage{algorithmic}
\usepackage{amsmath}
\usepackage{graphicx}
\usepackage{tabularx}
\usepackage{multirow}
\usepackage{moreverb}
\usepackage{rotating}
\usepackage{tcolorbox}
\usepackage{dialogue}
\usepackage{float}
\usepackage{cleveref}
\usepackage{footnote}
\usepackage{mathrsfs}
\usepackage{booktabs}
\usepackage[htt]{hyphenat}
\usepackage{dblfloatfix}
\usepackage{fancyhdr}

\fancypagestyle{specialfooter}{
  \fancyhf{}
  
  \fancyfoot[L]{\footnotesize DOI 10.1109/TSE.2020.3040935 \textcopyright  2020 IEEE. Personal use of this material is permitted. Permission from IEEE must be
obtained for all other uses, in any current or future media, including
reprinting/republishing this material for advertising or promotional purposes, creating new
collective works, for resale or redistribution to servers or lists, or reuse of any copyrighted
component of this work in other works.}
}
\newfloat{excerpt}{h}{lob}
\floatname{excerpt}{Excerpt}
\Crefname{excerpt}{Excerpt}{Excerpts}
\newcolumntype{Y}{>{\centering\arraybackslash}X}

\begin{document}

\title{A Wizard of Oz Study Simulating API Usage Dialogues with a Virtual Assistant}

\author{Zachary~Eberhart,~Aakash~Bansal,~and~Collin~McMillan%
\IEEEcompsocitemizethanks{
  \IEEEcompsocthanksitem The authors are with the Department
  of  Computer Science and Engineering, University of Notre Dame, IN 46556.\protect\\
  E-mail: zeberhar, abansal1, cmc@nd.edu
  \IEEEcompsocthanksitem This paper has supplementary downloadable multimedia material available
  at https://github.com/ApizaCorpus/ApizaCorpus provided by the authors. This includes materials related to the experimental design, experimental results, and dialogue act annotations. This material is 1.3 MB in size.
}%
\thanks{Manuscript received ---- ----; revised ---- ----; accepted ---- ----. This work is supported in part by the NSF CCF-1452959 and CCF-1717607 grants. Any opinions, findings, and conclusions expressed herein are the authors’ and do not necessarily reflect those of the sponsors.}
}

\markboth{IEEE Transactions on Software Engineering,~Vol.~X, No.~X, November~2020}%
{Eberhart \MakeLowercase{\textit{et al.}}: A Wizard of Oz Study Simulating API Usage Dialogues with a Virtual Assistant}

\IEEEtitleabstractindextext{%
\begin{abstract}
Virtual Assistant technology is rapidly proliferating to improve productivity in a variety of tasks.  While several virtual assistants for everyday tasks are well-known (e.g., Siri, Cortana, Alexa), assistants for specialty tasks such as software engineering are rarer.  One key reason software engineering assistants are rare is that very few experimental datasets are available and suitable for training the AI that is the bedrock of current virtual assistants.  In this paper, we present a set of Wizard of Oz experiments that we designed to build a dataset for creating a virtual assistant.  Our target is a hypothetical virtual assistant for helping programmers use APIs.  In our experiments, we recruited 30 professional programmers to complete programming tasks using two APIs.  The programmers interacted with a simulated virtual assistant for help -- the programmers were not aware that the assistant was actually operated by human experts.  We then annotated the dialogue acts in the corpus along four dimensions: illocutionary intent, API information type(s), backward-facing function, and traceability to specific API components. We observed a diverse range of interactions that will facilitate the development of dialogue strategies for virtual assistants for API usage.
\end{abstract}

\begin{IEEEkeywords}
Intelligent agents, Discourse, Software/Software Engineering, Wizard of Oz (WoZ), Virtual Assistants.
\end{IEEEkeywords}}

\maketitle
\thispagestyle{specialfooter}

\IEEEraisesectionheading{\section{Introduction} \label{sec:intro}}

Virtual assistants are software systems that interact with human users via natural language and perform tasks at the request of those users~\cite{cooper2004personal}.  Virtual assistants for everyday tasks (e.g., Cortana, Alexa, Siri) are proliferating after a period of heavy investment -- a confluence of sufficient training data, advancements in artificial intelligence, and consumer demand have fed rapid growth~\cite{white2018skill}.

Many of the achievements of virtual assistants for everyday tasks are beginning to be brought to specialty applications such as medicine~\cite{brown2017conversational} and education~\cite{lv2015virtual}.  However, a key observation is that these applications are quite specific (e.g., not education in general, but a specific type of geography for a specific age group of students).  The reason is that data collected for one application is difficult to generalize to other applications -- a virtual assistant must master both the language and the strategies humans use to move through conversations, and this problem is simply too complex for existing AI technologies to learn without detailed, specific training data~\cite{he2018decoupling}.  A relevant dataset must be collected and annotated for every type of conversation in which a virtual assistant needs to converse.  For example, a virtual assistant for everyday tasks would require different training data to recommend a restaurant and to reserve a table at that restaurant~\cite{whittaker2002fish, schmidt2018industrial}.

Virtual assistants for software engineering tasks suffer from the same hunger for data.  Despite long-recognized demand for virtual assistants to help programmers~\cite{boehm2006view, robillard2017demand}, working relevant virtual assistant technology remains something of a ``holy grail''.  Several research prototypes have made significant advances (see \Cref{sec:va4se}), but a major barrier to progress is a lack of well-understood, annotated datasets that are specific to software engineering tasks.  A survey in 2015 by Serban~\emph{et al.}~\cite{serban2015survey} found none related to SE tasks, and since that time only one has been published to our knowledge, targeting the task of bug repair~\cite{wood2018detecting}.  

One reason for the lack of suitable datasets is the investment cost necessary for experiments in numerous target tasks, and a perceived disincentive in terms of publication versus data and software artifacts~\cite{Howison:2011:SSP:1958824.1958904}.  A recent book by Rieser and Lemon~\cite{rieser2011reinforcement} provides clear guidance regarding how to build dialogue systems for virtual assistants, with particular focus on the design of experiments for data collection.  A major theme of the book is that, despite a perception that data collection experiments yield few immediate research outcomes, in fact the experiments provide answers to research questions about how people seek knowledge to perform tasks.  These answers are critical to the later design of virtual assistants, in addition to the data produced.  To this end, Rieser and Lemon establish two first steps towards building a virtual assistant: 1) conduct ``Wizard of Oz'' experiments to collect simulated conversation data, and 2) annotate every utterance in the conversations with dialogue act types.

A Wizard of Oz experiment is one in which a virtual agent is simulated.  Participants interact with a virtual assistant to complete a task, but they are unaware that the virtual assistant is actually operated by a human ``wizard''.  The deception is necessary because people communicate differently with machines than they do with other humans~\cite{dahlback1993wizard} and our objective is to create data for a machine to learn strategies to converse with humans.  The key element of these strategies are ``dialogue acts'': a dialogue act is a spoken or written utterance that accomplishes a goal in a conversation~\cite{bach1979linguistic}.  A conversation is composed of a series of utterances taken by different speakers, and each utterance functions as a dialogue act.  For example, the utterance ``tell me how to find my classroom'' is a dialogue act explicitly requesting information, whereas ``this is the wrong classroom'' is a statement that, depending on context, may imply a request for information.  A virtual assistant must be able to recognize when a human is, e.g., requesting information and generate an appropriate response, and to do that it relies on training data in which humans have annotated the dialogue acts in conversations.

At the same time, in software engineering, one task that cries out for help from virtual assistants is API usage: programmers trying to use an unfamiliar API to build a new software program. The authors of APIs are often not available to answer questions, and web support (e.g., StackOverflow) is neither a guarantee nor immediately available, which makes the rapid help a virtual assistant can provide more valuable. As Robillard~\emph{et al.}~\cite{robillard2017demand} point out, API usage is a high value target for virtual assistants due to the high complexity of the task, and a tendency to need the same information about different APIs~\cite{robillard2009makes, duala2012asking}.  In other words, programmers often have the similar kinds of questions about different APIs (e.g., ``Is there an API type that provides a given functionality?'' or ``How do I determine the outcome of a method call?''), even if the tasks performed by the APIs are not similar -- the similarity of questions makes API usage a good target for virtual assistants, since the virtual assistants are likely to be able to learn what programmers need to know.  

There are many scenarios in which programmers could benefit having a virtual assistant to assist with API usage.  In some cases, programmers may have limited access to traditional documentation; for instance, blind programmers rely heavily on API documentation to understand the structure of code~\cite{mealin2012exploratory}, but they are limited to existing screen-reading tools. These programmers may prefer to query a virtual assistant for specific information, rather than navigate large documents. Other user types, like children or novice programmers may find a virtual assistant more approachable than traditional documentation. A virtual assistant may also improve the usability of IDEs for mobile platforms with limited screen space (e.g., smartphones) by enabling programmers to find targeted answers to API questions without navigating away from the editor. 

In this paper, we conduct Wizard of Oz experiments designed to lay a foundation for the creation of virtual assistants for API usage.  We hired 30 professional programmers to complete programming tasks with the help of a ``virtual assistant,'' which was operated by a human wizard. The tasks involved designing a program that met specified objectives using an API (described in \Cref{sec:tasks}).  The programmers conversed with the virtual assistant, though they were not aware that it was operated by a human.  Each programming session lasted approximately 90 minutes.

We then annotated the dialogue acts in all 30 conversations along four dimensions. We labeled \textit{illocutionary} dialogue act types by adapting the dialogue act annotation scheme from the AMI conversation corpus~\cite{mccowan2005ami}, which consists of 14 coarse-grained illocutionary types extracted from simulated business meetings. We annotated domain-specific \textit{API} dialogue act types by adapting a taxonomy of API information types by Maalej and Robillard~\cite{maalej2013patterns}, which consists of 12 labels corresponding to domains of API knowledge. We annotated the \textit{backward-facing} function of each dialogue act by adapting a set of labels provided by AMI scheme to describe the relationships between utterances. Finally, we annotated specific API components (i.e., methods, structs, and variable names) that were referenced in each utterance in order to observe the \textit{traceability} between specific concepts and the language used to discuss them. This multi-faceted characterization of the corpus will enable downstream tasks in the development of a virtual assistant for API usage.

This paper makes the following specific contributions to the field of software engineering:

\smallskip
\begin{enumerate}
\item A corpus of 30 Wizard of Oz dialogues, comprising 44 hours of programming activity and including two separate APIs.
\item The results of those programming sessions, including programmers' comments, ratings of the simulated virtual assistant, and performance on the task sets.
\item Corpus annotations along 4 dimensions.
\item Several recommendations and considerations for future virtual assistant developers.
\end{enumerate}

\section{Background and Related Work}
\label{sec:background}

This section discusses background on the problem we target, supporting technologies, and related work.

\subsection{Problem Statement and Scope}
\label{sec:prob}

The problem we target in this paper is that the composition and patterns of dialogue acts are not known for conversations between programmers and virtual assistants during API usage tasks.  This problem is significant because information about these dialogue acts must be known in order to create lifelike, efficient dialogue strategies for virtual assistants.  The situation is a ``chicken or egg'' question because in order to obtain the dialogue act structure, one must have conversations between programmers and virtual assistants, but to have a virtual assistant one must know the dialogue act structure.  Software engineering literature does not describe dialogue acts for API usage conversations, which impedes development of usable virtual assistants.

To that end, the scope of this paper encompasses the following activities: 1) conducting Wizard of Oz experiments designed to promote programmer-wizard interactions related to API usage, 2) annotating the conversations with multi-dimensional dialogue act types that are likely to be useful for many downstream tasks, and 3) discussing the key insights from these experiments and annotations that will guide future development of dialogue strategies for virtual assistants for API usage. 

This scope is already quite extensive, so we note that we do not yet attempt to create a working model of dialogue strategy (such as one based on reinforcement learning). To train and validate such a model, researchers must create a simulated learning environment, compare various policy-training approaches, and evaluate the system~\cite{rieser2011reinforcement}. We also do not attempt to perform statistical analysis of factors affecting the performance of participants in the experiments, as our experimental design promotes variety in wizard-programmer interactions over carefully balanced treatments, as described in the following section.



\subsection{Wizard of Oz Experiments}
\label{sec:woz}

A \textbf{Wizard of Oz experiment} is one in which a human (the user) interacts with a computer interface that the human believes is automated, but is in fact operated by another person (the wizard)~\cite{dahlback1993wizard}.  The purpose of a Wizard of Oz experiment is to collect conversation data unbiased by the niceties of human interaction; people interact with machines differently than they do with other people~\cite{dahlback1993wizard}.  These unbiased conversation data are invaluable for kickstarting the process of building an interactive dialogue system, as they ``deliver a more or less complete specification of the system's input/output behaviour''~\cite{bernsen1994wizard}. We direct readers to a comprehensive survey by Riek~\emph{et al.}~\cite{riek2012wizard} for further justification and examples of Wizard of Oz experiments.

In a Wizard of Oz experiment, researchers must provide the participants with some specific \textbf{scenario}. A scenario is ``a task to solve whose solution requires the use of the system, but where there does not exist one single correct answer''~\cite{dahlback1993wizard}. A well-designed scenario promotes interaction relevant to the simulated system~\cite{1374824}. To that end, it is important for the scenario to place constraints on both the user and the wizard. Such constraints may include limitations on the resources available to the user, or the types of responses the wizard can generate.  Researchers often provide the wizard with an interface that simplifies and expedites the process of generating a response~\cite{dow2005wizard}.

Rieser and Lemon provide an excellent summary of the state-of-the-art in virtual assistant development in their recent book~\cite{rieser2011reinforcement}.  In short, they explain that a highly-effective method to kick start development of virtual assistants is to conduct Wizard of Oz experiments to collect conversations between humans and simulated virtual assistants (wizards).  The data from those experiments can then be used to design a virtual assistant prototype via reinforcement learning (see \Cref{sec:strats}).  Later, as more people interact with the virtual assistant, real-world data can be collected. The process boils down to collecting Wizard of Oz data and annotating the dialogue acts in that data. 

As Rieser and Lemon explain, Wizard of Oz experiments designed to support dialogue system development must strike a balance between promoting \emph{realistic} interactions (i.e. ones that cleanly demonstrate common patterns and behavior) and covering a \emph{diverse range} of interactions (i.e. ones that demonstrate behavior in less-common circumstances). To that end, experiments should be designed to target a narrow task domain while not imposing extraneous restrictions on participant behavior, in order to explore a broad range of ``intuitive'' strategies employed by participants~\cite{rieser2011reinforcement}.

Data from Wizard of Oz experiments have a variety of direct applications in the development of a virtual agent; for instance, they can be used to improve a system's natural language understanding capabilities. Many modern natural language understanding frameworks (such as Alexa Skills~\cite{atefi2020examining} or Xatkit~\cite{daniel2020xatkit}) require explicit samples of user phrases that correspond to different dialogue acts; rather than attempt to intuit the phrases that real users would use, virtual assistant developers can extract real examples directly from the Wizard of Oz data. Other approaches to dialogue act classification use Wizard of Oz data to train statistical models to generalize to unseen inputs~\cite{wood2018detecting,okur2019natural}. That said, the key advantage to the Wizard of Oz approach described by Rieser and Lemon is that it enables researchers to efficiently design optimal \emph{dialogue strategies} for tasks in domains where no prior data is available.

\subsection{Dialogue Strategies}
\label{sec:strats}

A \textbf{dialogue strategy} is the decision-making process that a dialogue system follows at each step in a dialogue to determine what to say next~\cite{scheffler2002automatic}. For instance, when a user asks a question, a virtual assistant must decide whether to immediately respond with an answer, or to first elicit additional information from the user. A key observation is that there is a difference between the \emph{strategy} involved in a conversation and the \emph{language} used to implement that strategy~\cite{he2018decoupling}.  The language is expressed by the actual words used to render an utterance, while the strategy is expressed by the sequence of dialogue acts used.  E.g., the language ``A: Hello.  B: Hi.  A: Where should we eat?  B: At Joe's.'' versus the conversation flow/strategy: greeting, greeting, suggestion-elicitation, suggestion.

In dialogue systems, complex strategies do not only rely on the previous dialogue act; instead, they consider a broader \textbf{dialogue state} (that is, the dialogue system's internal representation of the dialogue), as well as any external knowledge available to the virtual assistant, to determine an appropriate response type. For an API usage task, a virtual assistant's internal dialogue state may include information about API components the user has previously mentioned, and its external knowledge would comprise information about available API resources. Developers of virtual assistants design dialogue strategies by defining what actions the system will take given some dialogue state. These strategies may be manually encoded or learned as statistical models.

Without prior conversational data, developers would have no empirical basis from which to design dialogue strategies for virtual assistants, and they would be forced to rely on intuition and trial-and-error. Even with relevant conversational data from a Wizard of Oz study, strategies based on the behavior exhibited by wizards are suboptimal, as wizards themselves are not expected to follow optimal dialogue strategies. Unlike true virtual assistants, wizards are unable to rapidly parse large knowledge databases, and instead rely on available search tools and mental models to determine what information may be valuable at a given point in a dialogue. Different wizards should be expected to use different dialogue strategies, informed by their domain-specific expertise, interpersonal skills, and communication styles. Even an individual wizard may experiment with multiple dialogue strategies as he/she observes which ones are more or less effective. Furthermore, methods that extract strategies directly from Wizard of Oz corpora may suffer from data sparsity, and are prone to overfitting~\cite{rieser2011reinforcement}.

Instead, Rieser and Lemon~\cite{rieser2011reinforcement} and Williams and Young~\cite{williams2003using} demonstrate how the data from Wizard of Oz experiments can be used to design a \textbf{reinforcement learning} problem in which a virtual agent automatically discovers optimal dialogue strategies.  In this reinforcement learning approach, the agent repeatedly engages in ``dialogues'' with simulated users, in which each dialogue turn is expressed as a descriptive dialogue act representation. Through these repeated interactions, the virtual agent is able to experiment with different dialogue strategies and discover the optimal action to take given any arbitrary dialogue state. 

In order to model this reinforcement learning problem, four distinct elements must be defined: an \emph{action space}, consisting of all dialogue acts available to the user and the system; a \emph{state space}, accounting for any relevant dialogue- and task-specific features; a simulated \emph{learning environment}, which updates the dialogue state after each system action; and a \emph{reward function}, which provides the system with feedback upon taking a certain action in a certain state. Annotated conversational data enable developers and researchers to design reinforcement learning models for dialogue strategies by  providing an empirical basis for deciding on the appropriate action and state spaces, determining the reward function, and creating a simulated learning environment that reflects the goals and behaviors of real users.


\subsection{Dialogue Acts}
\label{sec:das}

A \textbf{dialogue act} is a spoken or written utterance that accomplishes a specific purpose in a conversation. Dialogue act classification refers to the task of labeling utterances with descriptive dialogue act \emph{types}, such as ``greeting'' or ``information-elicitation''~\cite{reithinger1997dialogue}.

To create a dialogue system, both language and dialogue act types must be known for utterances in example conversations~\cite{rieser2011reinforcement}.  A virtual assistant must learn to mimic good strategies in terms of dialogue act flow (e.g., it must recognize that it should respond to a suggestion-elicitation with a suggestion).  Once it knows that it should respond with a particular dialogue act type (e.g., a restaurant-suggestion), it must then collect the information to portray in an utterance (e.g., a restaurant to recommend) and then convert the information into an understandable utterance in natural language.  As Serban~\emph{et. al}~\cite{serban2015survey} point out in a recent survey, several datasets, especially involving Wizard of Oz experiments, have been created for a variety of domains to serve as starting points for training virtual assistants.

It is important to note that a single dialogue act can serve multiple functions simultaneously; for instance, the utterance ``I have to work tonight'' in response to an invitation serves as an informative act as well as a rejection of the previous utterance. To account for these independent functions, annotators often label dialogue acts across several different \textbf{dimensions}~\cite{popescu2005dialogue}. 

Dialogue act dimensions can be broadly characterized as either communication-oriented or domain-oriented. Communication-oriented dimensions relate solely to the communicative role of an utterance in a conversation, rather than the content of the utterance. For instance, a communication-oriented dimension may capture the \emph{illocutionary} function of an utterance (i.e. the speaker's intention in producing an utterance) using dialogue act types such as ``inform'', ``suggest'', or ``offer''~\cite{popescu2005dialogue,mccowan2005ami}. A different communication-oriented dimension may describe the \emph{backward-facing} function of an utterance (i.e. how an utterance relates to a previous utterance in the conversation) with dialogue act types like ``accept'', ``repeat'', or ``answer''~\cite{Core1997CodingDW}. Other potential communication-oriented dimensions address features like time-management (e.g., ``stalling'') or discourse-management (e.g., ``change-topic'')~\cite{ISOStandard2017, DIT}. Communication-oriented dimensions and their associated dialogue act types describe conversational features that are largely domain-independent.

By contrast, domain-oriented dimensions relate to the specific subject matter of a conversation. Domain-oriented dimensions may identify the topic of an utterance, or the task that the speaker is executing~\cite{walker2001quantitative,leech2003generic}. Other dimensions may identify specific pieces of information that are communicated in an utterance, such as specific dates or locations; these narrow dimensions are often referred to as the ``slots'' or ``arguments'' of a dialogue act~\cite{young2007cued,frames}. 

Annotation schemes comprise one or more dimensions, each associated with a specific set of labels that may be used to characterize an utterance along that dimension. Most popular general-purpose dialogue act annotation schemes provide anywhere from 3 to 10 different dimensions (as in \cite{Core1997CodingDW, ISOStandard2017, DIT}), but researchers typically select or create an annotation scheme to suit to their particular research goals~\cite{clark2004multi}. 

Multi-dimensional annotation schemes can be flattened to one-dimensional schemes for use in downstream tasks. Rieser and Lemon~\cite{rieser2011reinforcement} explain how a multi-dimensional annotation scheme can be reduced to a one-dimensional action set to simplify dialogue strategy design for dialogue systems. In short, frequent combinations of dialogue act types in a corpus can be identified and associated with high-level, task-specific actions that map onto desired functionalities in a virtual assistant. However, if the original annotation scheme does not provide enough granularity to describe certain types of interactions or functionality, annotation along additional dimensions would be required.

\subsection{API Learning Resources}
\label{sec:resources}

Programmers use a variety of resources when learning and using APIs~\cite{gao2020exploring}. Typically, developers provide official documentation to accompany APIs, such as API reference documentation, tutorials, and example projects~\cite{meng2018application}. These resources are critical to API usability, as they describe the intended behavior of API components, specify constraints, suggest useful design patterns, and provide other ancillary information to facilitate API usage~\cite{maalej2013patterns}. In addition to official API documentation, programmers also make use of unofficial API resources available online, such as QA websites (e.g., StackOverflow), blogs, and code repositories~\cite{gao2020exploring}. These crowdsourced resources can provide broad coverage~\cite{parnin2011measuring}, but are of lesser value for  private APIs or obscure concepts~\cite{jiau2012facing}. When browsing API documentation, programmers may alternate between ``opportunistic'' approaches, which involve searching for resources to address problems as they arise, and ``systematic'' approaches, which involve building up an understanding of the overarching API functionality and design before implementing API components~\cite{meng2019developers}.

Unfortunately, APIs are often poorly documented in practice, as writing and maintaining documentation can be a costly and time-consuming endeavor~\cite{zhong2013detecting}. As a result, programmers often encounter challenges when learning and using APIs. Robillard and DeLine~\cite{robillard2011field} performed a large-scale study of API learning obstacles that affect professional developers. They found that the most severe obstacles were related to API documentation; specifically, inadequate intent documentation, pattern documentation, code examples, penetrability, and presentation were identified as the leading causes of difficulty in API learning. Similar studies reported on the issues most frequently raised by programmers when navigating API documentation~\cite{robillard2009makes,ko2011role}. These problems are broadly classified into two categories: problems with content (e.g., documentation that is incomplete, incorrect, ambiguous, or fails to link specific API functionality with high-level concepts) and problems with presentation (e.g., documentation that is bloated, fragmented, or contains excessive structural information). These problems can hamper the usability of an API~\cite{piccioni2013empirical}, leading to its misuse~\cite{zhang2018code,amann2018systematic}, or causing developers to abandon it altogether~\cite{robillard2011field}.

To help programmers overcome the challenges caused by inadequate API learning resources, researchers have investigated tools and techniques to assist with API navigation and improve usability. Some proposed tools improve upon traditional documentation search functionality by incorporating additional filtering and sorting options~\cite{de2013multi,treude2015tasknav}, or implementing alternative query similarity metrics~\cite{zhong2009mapo,ye2016word}. Several tools use online resources to augment official API documentation; for instance, the JADEITE~\cite{stylos2009jadeite} and Apatite~\cite{eisenberg2010apatite} tools count the number of web search results for various API components and highlight those that appear to be more popular, while other tools aim to enhance official documentation with key insights~\cite{treude2016augmenting}, FAQs~\cite{chen2014asked}, and up-to-date code examples~\cite{subramanian2014live} from the internet. Other researchers have focused their efforts on automatic API redocumentation, including techniques for automatic code summarization ~\cite{leclair2019neural} and API usage example generation~\cite{gu2016deep,buse2012synthesizing}. More recently, researchers have started looking into virtual agent technology to assist programmers with API usage.

\subsection{Virtual Agents in Software Engineering}
\label{sec:va4se}

Related work regarding virtual agents in the software engineering (SE) literature can be broadly categorized as either supporting experimentation or prototype virtual agents.  In terms of supporting experimentation, recent work at FSE'18~\cite{wood2018detecting} is the most similar to this paper.  In that work, the authors conducted Wizard of Oz experiments for debugging tasks and built an automated classifier for dialogue act types.  However, the one-dimensional dialogue act annotations in that study were rather general, such as ``statement'' or ``apiQuestion.''  This paper is different in that we have entirely new Wizard of Oz experiments for API usage and provide more thorough analysis by annotating additional dialogue act dimensions. 

Prototype virtual agents include APIBot~\cite{tian2017apibot} (a QA system for API documentation), WhyLine~\cite{ko2004designing} (a natural language debugging tool), TiQi~\cite{pruski2015tiqi} (a natural language interface to query software projects), and Devy~\cite{bradley2018context} (a virtual assistant that performs Git operations).  A comprehensive survey was recently conducted by Arnaoudova~\emph{et. al}~\cite{arnaoudova2015use}.  Our work most closely relates to APIBot~\cite{tian2017apibot}, a question-answering system for APIs, and OpenAPI Bot~\cite{edopenapi}, a chatbot to help users navigate REST APIs. However, there are several key differences: for instance, APIBot is not designed for multi-turn dialogues, and cannot request additional information from the user or consider dialogue history. OpenAPI Bot maintains some dialogue state information, but it does not proactively elicit information from the user. Furthermore, neither paper performs a user study to investigate what dialogue acts and strategies would be most valuable in a virtual assistant. Therefore, we view our work as complementing and enhancing existing work on virtual assistants in SE.


This work follows a history of empirical studies in software engineering~\cite{799955}. In their ``roadmap'' to empirical studies in SE, Perry~\emph{et. al}~\cite{Perry:2000:ESS:336512.336586} emphasize that ``empirical studies can be used not only retrospectively to validate ideas after they've been created, but also proactively to direct our research.'' Indeed, exploratory studies play an important role in motivating, guiding, and informing future work~\cite{1027796} (such as an exploratory study on feature location processes by Wang et al.~\cite{6080788} that directly inspired subsequent improvements~\cite{wang2013improving}). It is our hope that the present study will similarly facilitate the development of virtual assistant technology for SE.

\section{Wizard of Oz Experiments}
\label{sec:experiment}

This section describes the Wizard of Oz experiments we designed to simulate the experience of using an API with the help of a virtual assistant. We designed two scenarios in which programmers were asked to complete programming tasks using an API for an unfamiliar C library. The first scenario used the libssh networking library, while the second used the Allegro multimedia library. In lieu of documentation, we introduced the programmers to an ``experimental virtual assistant'' named \textit{Apiza}. Unbeknownst to the programmers, Apiza was controlled by a human (the  ``wizard''). 

The overarching rationale for this experimental design was to give rise to circumstances in which a variety of programmer-wizard interactions would be observed, in service of downstream research tasks. To this end, we made a number of decisions regarding the experimental design that restricted the scope of the study, as in the API selection and task design, as well as decisions to leave certain variables uncontrolled, as in the participant selection and assignment. These decisions are detailed in the remainder of this section.

\subsection{Participants}
\label{sec:part}

We distinguish between two participant roles in the experiments: the \emph{programmers} and the \emph{wizards}. No participant served as both a programmer and a wizard. All participants were asked to fill out an entry survey describing their backgrounds and levels of programming experience (summaries of these surveys are given in Appendix A). 

\subsubsection{Programmers}

We recruited 30 participants to serve as programmers. We recruited 2 locally through our university's Computer Science graduate program, 7 through various freelancer Subreddits, and the remaining 21 through the freelancing website Upwork. All programmers had experience using C in an academic or professional software engineering context and had no relevant experience with the API that they were to use in their scenario.

Each programmer participated in a single session. We gave half of the programmers the libssh scenario, and the other half the Allegro scenario. The two locally-recruited programmers participated in an on-campus office room using a laptop we provided. The rest of the programmers worked remotely in the environment of their choice, using their own computers.  All programmers worked in a virtual machine running Ubuntu 16.04. We asked programmers to work from this virtual environment for two reasons: to ensure that all of the necessary libraries and compilation tools were properly installed and configured, and to reduce the number of potential distractions on the user's screen.

\subsubsection{Wizards}
\label{sec:wiz}
We recruited 6 participants to serve as wizards. The first author served as the wizard for ten sessions. The other wizards -- 2 computer science graduate students and 3 professional software engineers -- served for between one and six sessions each. Qualifications for wizards were identical to the programmers: they had used C in an academic or professional software engineering context, and had not previously been exposed to the APIs in our experiment. One wizard only participated in the libssh scenario, one only participated in the Allegro scenario, and four participated in both the libssh and Allegro scenarios. All wizards worked remotely and used their own computers.

\begin{figure}[b]
  \includegraphics[width=\linewidth]{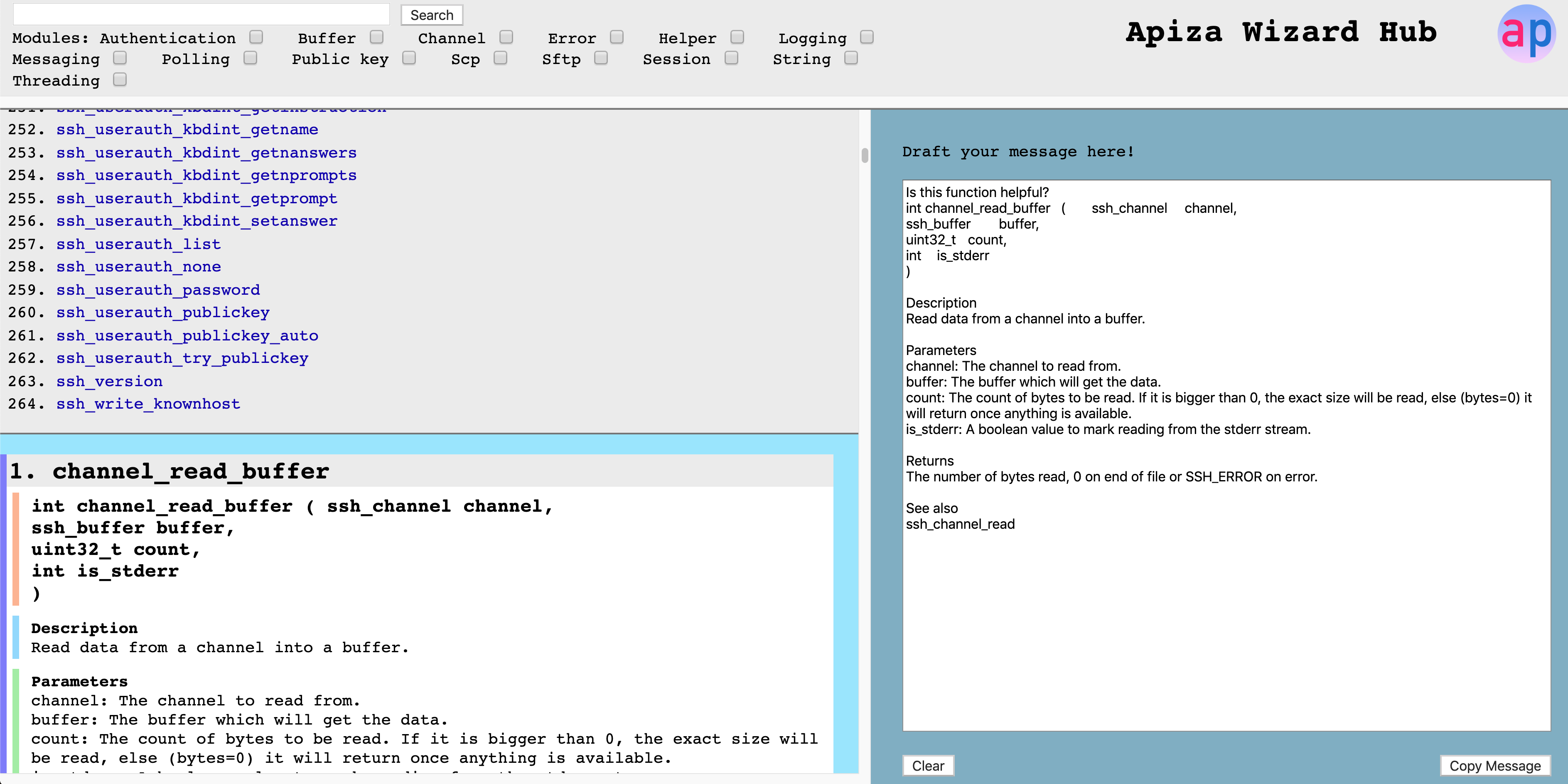}
   \caption{Tool used by wizards in the Wizard of Oz experiments. The header bar allowed wizards to search for components by keyword and filter by categories defined in the API documentation. The left half of the interface displayed a list of the names of all components that satisfied the search, followed by the complete documentation for each of those components. The right half contained a text box used to draft messages.}
  \label{fig:tool}
\end{figure}

In line with other Wizard of Oz experiments, we provided all wizards with a custom tool to draft messages and navigate documentation (shown in \Cref{fig:tool}). The tool allowed wizards to search for API components by keyword and category. Wizards could click on labeled sections of function documentation (e.g., Description, Returns, Parameters) to copy individual sections, or click on a function name to copy the entire documentation for that function. The tool also provided searchable links to the header files. All wizards (excluding the author) underwent a brief training session prior to their first session, in which they were introduced to the purpose and parameters of the study, given examples of interactions from pilot studies, and asked to spend no more than 30 minutes familiarizing themselves with the API and experimental tool.

A feature of our study, differing from Wood~\emph{et al.}~\cite{wood2018detecting} but similar to Benzmüller~\emph{et al.}~\cite{ullerwizard} and Kruijff-Korbayov{\'a}~\emph{et al.}~\cite{kruijff2006sammie}, is that we hired wizards as experimental participants in addition to the programmers.  The decision to hire multiple wizards was intended to enable us to collect a more diverse set of dialogue strategies; by hiring wizards who were unfamiliar with the APIs and allowing them to participate in multiple sessions, we anticipated that there would be a learning effect in which wizards would adopt new strategies as they became more familiar with the API upon completing successive sessions.

\subsection{Scenarios}
\label{sec:tasks}

\begin{table*}[t]
	\centering
	\caption{Summaries of the two experimental scenarios.} 
	\begin{tabular}{ccccccc}
		\toprule
		API & Domain & \# Participants & \# Tasks & Task Detail & Completed in Order & API Examples in Skeleton Code\\ 
		\midrule
		libssh & Networking & 15 & 5 & Direct instructions & True & False\\
		Allegro & Multimedia & 15 & 7 & Open-ended & False & True \\
		\bottomrule
	\end{tabular}
	\label{tab:tasks}
\end{table*}

We created two scenarios, which consisted of sets of software design tasks based on APIs for two different C libraries: the \emph{libssh} API, and the \emph{Allegro} API. 

We chose to use C APIs because they consist of basic data structures, variables, and functions; APIs in object-oriented languages (e.g., Java) often contain more complex class hierarchies, which could result in many programmer-wizard interactions that may be irrelevant to other APIs (e.g., many web APIs). We chose APIs from two different domains (networking and multimedia) to observe a broader range of interactions than might be observed using a single API. We chose the Allegro and libssh APIs in particular because they are both fairly large (with several hundred public functions and data structures) and well-documented. Summaries of both scenarios are shown in \Cref{tab:tasks}.

Prior to the experiment, the first author completed the tasks in each scenario to ensure that they  were possible to complete and of reasonable difficulty. We chose software design tasks (in contrast to, e.g., software maintenance tasks) to promote interaction between the programmers and wizards by withholding information about API functionality in the task specifications.

We prepared both scenarios ahead of time in the virtual machine used by the programmers. For each scenario, programmers only needed to edit a single file containing some skeleton code. After making changes, programmers were responsible for running a premade \texttt{make} file to observe the program's behavior and evaluate their progress. The task descriptions and skeleton code are available in our online Appendix (see \Cref{sec:conclusion}). They appear as \texttt{task\_description.pdf} and \texttt{skeleton\_code.c} in the \texttt{allegro/} and \texttt{libssh/} folders.

The programmers were allowed to use any basic text editor inside the virtual machine to complete the tasks. While this decision introduced some variability, it meant that programmers did not need to waste time adapting to a totally unfamiliar development environment. It also meant that programmers would not be able to take advantage of features available in some IDEs that reveal additional API information (e.g., by indicating return or parameter types, or suggesting auto-completions), encouraging them to direct a broader range of questions to the wizards.

\subsubsection{Scenario 1: libssh}

The first scenario involved using the libssh API to programmatically create and employ SSH network connections. It consisted of five tasks, presented in order of increasing difficulty. The programmers were instructed to complete these tasks in order.

The first task simply directed the programmers to compile the program and observe its behavior. This task did not require use of the API, but we included it to ensure the programmers understood how to evaluate their progress. The second task directed programmers to create a new \texttt{ssh\_session} object. We intended for this task to be simple, requiring only a single API call (\texttt{ssh\_new}), in order for the programmers to become acquainted with Apiza.

We divided each of the remaining tasks into a series of subtasks. The wording of these subtasks generally hinted at relevant API components (e.g., a subtask directing programmers to connect to an ssh server could be completed using the function \texttt{ssh\_connect}). The third task directed the programmers to set up the \texttt{ssh\_session} created in the previous task by connecting to a server (the localhost), authenticating the server, and disconnecting from the server. The fourth task directed the programmers to complete an empty method called \texttt{show\_remote\_user} by creating and opening an \texttt{ssh\_channel}, executing the \texttt{who} command on the channel, reading the response, and shutting down the channel. This task used many methods similar to those previously used (e.g., \texttt{ssh\_channel\_new}), as well as new methods to execute on and read from a channel. The final task directed programmers to fill out an empty method called \texttt{sftp\_operations} by creating and initializing an \texttt{sftp\_session}, creating a new file in a new directory, writing a string to the file, and finally closing the file. 

\subsubsection{Scenario 2: Allegro}


The second scenario involved using the Allegro multimedia library to add features to a simple video game. There were a few key differences between the two scenarios. 

Whereas programmers in the libssh scenario were provided a nearly empty source file to work with, we provided Allegro programmers with skeleton code in which several features were already implemented (such as the display and core game loop). These features needed to be correctly implemented before any other interesting tasks could be completed, but they were too complicated and used too few API functions to serve as good tasks themselves.

After observing that some programmers in the libssh scenario finished all five tasks with time to spare, we decided to include seven tasks in this scenario. Additionally, we did not include the initial compilation and observation of program behavior as a separate task, as in the libssh scenario. We instructed the programmers to compile and observe the program before beginning the session, to allow them more time to work on the tasks that actually involved the API. 

Unlike in the libssh scenario, we allowed the programmers to work through tasks in any order to prevent them from getting stuck on any one problem (in practice, nearly all programmers worked through them in the order provided). Also unlike the libssh tasks, these tasks provided only high-level descriptions of the features that were to be incorporated and 2-3 details or hints. As such, these problems were a bit more open-ended than those in the libssh set. This decision was intended to procure a wider range of programmer-wizard interactions.

The first task directed the programmers to add keyboard functionality to the game. This required installing the keyboard subsystem, registering it as an event source, and checking for keyboard events. Though this task required the use of at least 4 API methods, they were analogous to methods already implemented in the template program (e.g., the \texttt{al\_get\_display\_event\_source} method may have hinted at the existence of the \texttt{al\_get\_keyboard\_event\_source} method).

The remaining tasks directed the programmers to add a ``game over'' sound effect, show a score on the display, draw images on the display, rotate the images appropriately, pause the game when the player clicked on the display, and make the display resizable. These tasks generally required similar steps, such as identifying the correct subsystems and handling events. The later tasks required the programmers to understand more complex aspects of the API.







\subsection{Methodology}

At the start of each session, we instructed the programmer to open the virtual machine testing environment and login to a Slack channel for communication. At the same time, we had the wizard participant login to the same Slack channel using an account named ``Apiza''.

In Slack, we provided the programmer with a document describing the scenario, including the list of specific tasks to complete with the unfamiliar API (as described in the previous section). We asked that all questions relating to the API be directed via Slack text messages to our ``experimental virtual assistant'' called Apiza. We explained that Apiza was an ``advanced AI,'' able to carry out ``organic, natural-language conversation'' and discuss ``both high-level and low-level functionality.'' 

Once the programmer confirmed that he or she understood the description and the tasks, we started a timer and instructed the programmer to begin. For the next 90 minutes, the programmer worked through as many of the tasks as he or she could. Throughout, the programmer sent messages to the wizard, who answered them as quickly and correctly as he or she could. We instructed the wizard that Apiza's responses didn't need to seem ``robotic,'' but at no point was the wizard to reveal that Apiza was a human.  

During the session, we did not allow the programmer to access the API's documentation. While this does not necessarily represent the most-likely use case for a virtual assistant for API usage, it is a constraint necessary in the vast majority of Wizard of Oz experiments to force programmers out of their habits and into using the experimental tool~\cite{riek2012wizard}. However, we did permit the programmer to search the internet for general programming questions (e.g., related to C syntax) in order to narrow the scope of the dialogue. 

Unlike the programmer, the wizard had access to several API resources that would be available to a hypothetical virtual assistant. The wizard was able to browse the API documentation and header files and search for keywords using the tool described in \Cref{sec:wiz}. Additionally, we permitted the wizard to search for API-related information on the internet, as several tools for API navigation use online resources to augment the official documentation (as described in \Cref{sec:resources}). The main factor limiting the wizard's access to information was the pressure to generate a timely response. 

When the time ran out or the programmer finished all of the tasks, we instructed the programmer to stop working. We then asked the programmer to send us his or her code and the list of all URLs visited in the course of the study.  We also asked them to complete an exit survey; as recommended by Reiser and Lemon~\cite{rieser2011reinforcement}, the exit survey included the {\small \texttt{PARADISE}}~\cite{hajdinjak2006paradise} questions, which rated user satisfaction on a 5-point Likert scale.  

\subsection{Data Collection}

We collected six key data items for every experimental session:

\begin{enumerate}
	\item The programmer's entry survey.
	\item The wizard's entry survey.
	\item The dialogue between the programmer and the wizard.
	\item The source code written by the programmer.
	\item A record of any websites the programmers visited during the session for general C syntax questions.
	\item The programmer's exit survey.
\end{enumerate}

Some programmers also offered additional comments about their experience with Apiza. All data are made available in our online Appendix (see \Cref{sec:conclusion}).

\section{Experiment Results}
\label{sec:experiment_results}

In this section, we present the results of our Wizard of Oz experiments. We outline the basic structure and descriptive statistics of the collected corpus, and briefly examine the programmers' task performance and satisfaction with the simulated system. Note that we do not draw statistical conclusions regarding the effects or interactions of experimental factors, as these were not within the scope of our study (see \Cref{sec:prob}). Nevertheless, we do make several key qualitative observations and discuss them in the broader experimental context.

\subsection{Dialogues}
\label{sec:dialogues}

\begin{table}[t]
	\caption{Comparison of our corpus to other Wizard of Oz corpora. The numbers of words and unique words in each corpus are shown where available.} 
	\centering
	\begin{tabularx}{\linewidth}{Xcccc}
		\toprule
		\multirow{2}{*}{Task Domain}&\# of&\# of&\# of &\# Unique\\
		&Dialogues&Turns&Words&Words\\
		\midrule
		APIs (this paper)&$30$&$1947$&$47928$&$3190$\\
		Debugging~\cite{wood2018detecting}&$30$&$2243$&$50514$&$4162$\\
		Design~\cite{pineda2002dime}&$31$&$3606$&$27459$&--\\
		General QA~\cite{bertomeu2006contextual}&$33$&$2534$&$125534$&--\\
		Audio player~\cite{rieser2005corpus}&$72$&$1772$&17076&--\\
		Tutoring~\cite{benzmuller2006corpus}&$37$&$1917$&$12346$&--\\
		Mission planning~\cite{katsakioris2019corpus}&$22$&$1738$&--&--\\
		\bottomrule
	\end{tabularx}
	\label{tab:wozcorpora}
\end{table}

We collected 30 API usage dialogues. This corpus is similar in size to published Wizard of Oz corpora across a broad range of domains, as shown in \Cref{tab:wozcorpora}.

In general, the programmers and wizards sent messages in turn. Often, these were question-answer pairs, with the programmer querying some functionality of the API and the wizard providing the answer. Other types of interactions occurred as well, such as greetings, assessments, and side-exchanges -- these are explored in greater detail in Sections \Cref{sec:annotation} and \ref{sec:annotation_results}. The following excerpt from Dialogue \#19 in the corpus typifies the interactions that occurred in the dialogues. Messages are labeled with ``PRO'' or ``WIZ'', denoting the speaker as a ``Programmer'' or ``Wizard.''

\begin{excerpt}
\begin{tcolorbox}[left=5pt,right=-15pt,top=5pt,bottom=5pt]
\begin{flushleft}
\begin{dialogue}
\speak{Pro} allegro keyboard input
\vspace{0.1cm}
\speak{Wiz} You can save the state of the keyboard specified at the time the function is called into the structure pointed to by \texttt{ret\_state}, using \texttt{al\_get\_keyboard\_state}
\vspace{0.1cm}
\speak{Pro} Whats the function signature for \texttt{al\_get\_keyboard\_state} 
\vspace{0.1cm}
\speak{Wiz} \texttt{void al\_get\_keyboard\_state( ALLEGRO\_KEYBOARD\_STATE *ret\_state)}
\end{dialogue}
\end{flushleft}
\end{tcolorbox}
\label{exc:basic}
\end{excerpt}

Across all dialogues, participants collectively generated 1927 Slack messages (also referred to as ``turns''). Wizards and programmers sent similar quantities of messages, averaging to 31.8 messages/dialogue sent by programmers and 33.1 messages/dialogue sent by wizards. The frequency was also similar across the two tasks; participants sent an average of 68.5 messages in the libssh scenario, compared to 61.3 sent in the Allegro scenario.

The dialogues contain a total of 47928 word tokens\footnote{Word tokens were generated using the \texttt{word\_tokenize} method from Python's \texttt{nltk.tokenize.punkt} module.} with a vocabulary size of 3190 words. Wizards used considerably more words (41185) and drew from a larger vocabulary (2988) than programmers, who used 6743 words and 880 unique words. This disparity in word usage is to be expected; programmers manually wrote the majority of their messages' content, and had access to only limited information. By contrast, wizards frequently copied and pasted large chunks of pre-written documentation for the user.

\subsection{Task Completion}
\label{sec:task_completion}

Not every programmer completed every task. This fact is valuable because the task completion rate can be used as a metric to characterize the ``success'' of individual dialogues, which may be useful in designing a reward function for a model for dialogue strategy development.

We calculate two specific metrics: the task \emph{attempt rate} (that is, the proportion of programmers that attempted a task) and the task \emph{completion rate} (of the programmers who attempted a task, the proportion who successfully completed it). We considered a programmer to have ``attempted'' a task if he or she wrote at least one line of code directly related to the task. If it was ambiguous whether a line of code related to a task, we referred to the dialogue for additional context. We considered a programmer to have ``completed'' a task if he or she correctly implemented API components necessary to satisfy that task's requirements, even if the code did not compile or execute\footnote{For instance, tasks in the libssh scenario that relied on an ssh connection established in prior tasks could be considered completed even if the prior tasks setting up that connection were unsuccessful. Similarly, tasks could be marked as completed even if syntax or control flow errors unrelated to the API usage for that particular task prevented code related to that task from  executing.}.

\Cref{tab:taskcompletion} shows programmers' attempt and completion rates for each task, as well as the total number of messages in the corpus that were related to task. For instance, $71.4\%$ of all programmers attempted Task 4 of the libssh scenario; of those programmers, $40.1\%$ completed the task; and in total, 455 messages sent by all participants in the libssh scenario related to Task 4.

\begin{savenotes}
	\begin{table}[t]
	\caption{Programmers' performance on the tasks. ``Attempt rate'' refers to the percentage of all programmers that wrote at least one line code directly related to the task. ``Completion rate'' refers to the percentage of those programmers that successfully completed the task. ``\# Messages'' refers to the total number of messages in the corpus related to the task.}
		\centering
		\begin{tabular}{lcrrr}
		\toprule
			Scenario & Task & Attempt Rate         & Completion Rate & \# Messages\\ 
		\midrule
			\multirow{5}{*}{libssh\footnote{Only 14 sessions are considered here, as one participant in the libssh scenario did not submit a source code file.}}&1&$100.0$&$100.0$&0\\
			&2&$100.0$&$78.6$&194\\
			&3&$100.0$&$21.4$&723\\
			&4&$71.4$&$40.1$&455\\
			&5&$51.1$&$28.0$&317\\
		\midrule
			\multirow{7}{*}{Allegro}&1&$100.0$&$66.7$&473\\
			&2&$93.3$&$78.6$&446\\
			&3&$80.0$&$41.7$&303\\
			&4&$40.0$&$66.8$&109\\
			&5&$6.7$&$100.0$&26\\
			&6&$6.7$&$100.0$&43\\
			&7&$0.0$&N/A&0\\
		\bottomrule
		\end{tabular}
	\label{tab:taskcompletion}
	\end{table}
\end{savenotes}{}

Attempt and completion rates for different tasks varied between $0\%$ and $100\%$.  We generally observed lower attempt and completion rates for the later tasks, which were more difficult and for which the programmers may have had less time, depending on their performance on earlier tasks. Tasks 5 and 6 of the Allegro scenario each have a success rate of $100\%$ because only one programmer attempted and completed those tasks.  Programmers generally finished one task before moving to the next; however, they occasionally moved on from a task without successfully completing it, including several programmers in the libssh scenario (despite instruction to complete the tasks in order). 

\subsection{User Satisfaction}
\label{sec:user_satisfaction}

\begin{figure}[t]
  \includegraphics[width=\linewidth]{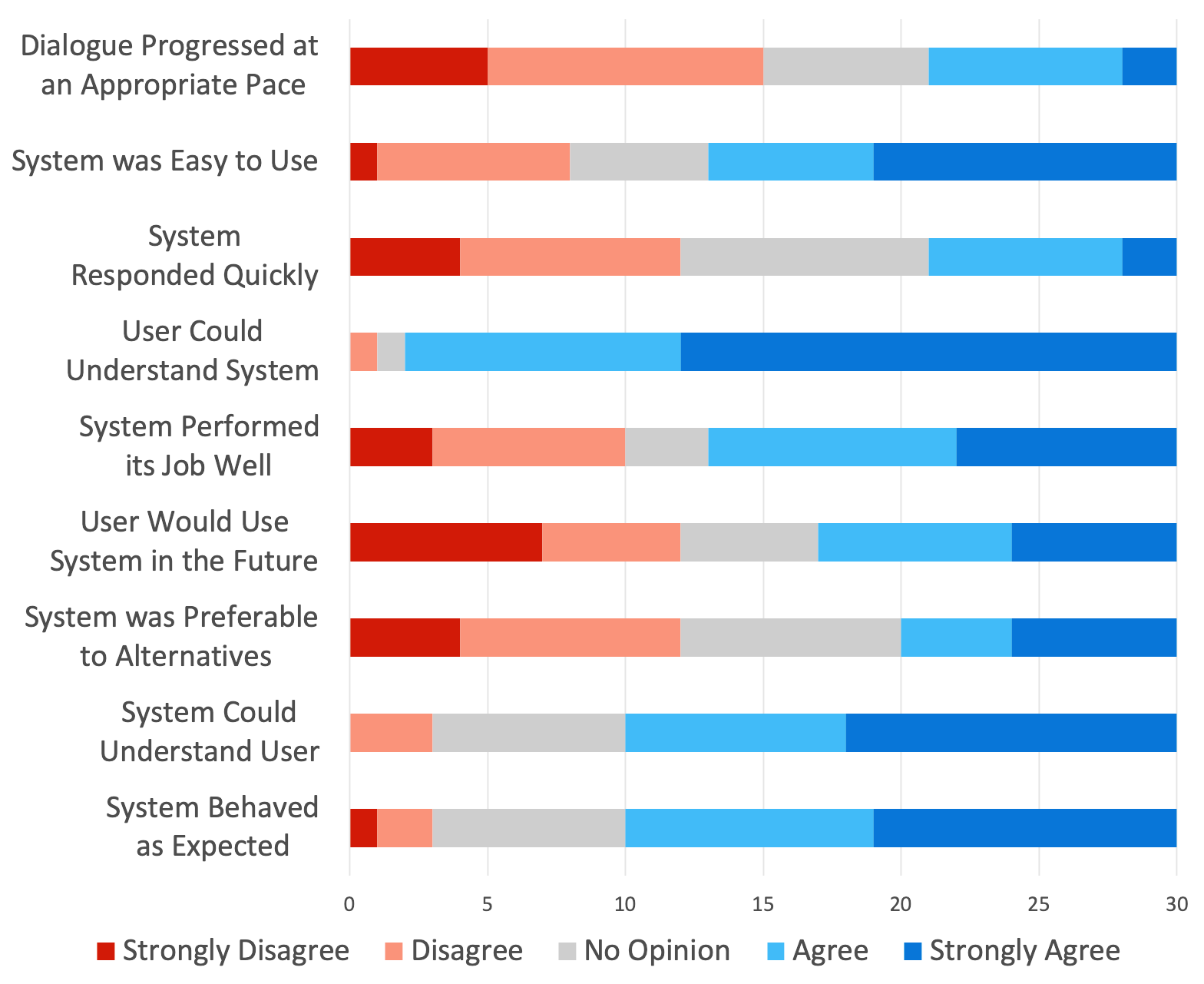}
	\caption{Programmers' PARADISE ratings for the Apiza system.}
  \label{fig:paradise}
\end{figure}

In addition to observing the programmers' objective performance on the tasks, we also asked programmers to fill out surveys subjectively rating their satisfaction with the ``virtual assistant'' system on a 5-point Likert scale. The survey questions were taken from the \texttt{PARADISE}~\cite{hajdinjak2006paradise} framework for automatic dialogue evaluation. To measure different dimensions of user satisfaction, the authors of the PARADISE framework created a set of 9 survey questions corresponding to features such as ``Interaction Pace'' and ``Task Ease.'' These questions have been widely adapted and used in subsequent dialogue studies~\cite{rieser2011reinforcement}. \Cref{fig:paradise} summarizes the results of the user survey. We include the full list of survey questions and programmer responses in our online Appendix (see \Cref{sec:conclusion}).

The features corresponding to the system's ability to understand user turns and the understandability of the system's turns were the most highly-rated, scoring around 4 points on average. The features related to the pace of the conversation  and the speed of the system's response received the lowest ratings, around 2.8 points on average. Scores varied greatly across raters; the harshest rater gave an average score of 1.5, while the most generous rater awarded an average score of 4.9. 

That the features relating to interaction pace generally scored relatively low and those related to system understanding scored high is not surprising given the nature of the Wizard of Oz deception. For users accustomed to real virtual assistants, the system's high intelligence came as a pleasant surprise, while its sluggishness was a source of frustration.  As one programmer wrote in a comment, ``Most of the time the first answer given by Apiza directly answered the question [...] The wait was a bit annoying though. I'm used to getting answers (well, search results) within a few seconds of entering a query.''

\subsection{Observations}
\label{sec:obs}

We report several qualitative observations regarding the results of the experiments.

\subsubsection{Programmer behavior} 

Programmers engaged in a broad variety of conversational behaviors related to API usage. The most prominent behavior exhibited by nearly all participants was asking ``How?'' questions. These questions were widely used by programmers to introduce task transitions and elicit high-level guidance (e.g.,  ``How do I play a sound in Allegro?'') but programmers also asked these ``How?'' questions in service of lower-level outcomes (e.g., ``How do I access the keycode from an event object''). Other frequently-observed behaviors included questions regarding the functionality of specific API components (e.g., ``What is the size parameter in al\_load\_font''), the structure and contents of the API (e.g., ``list all ssh connection functions''), requests for example code (e.g., ``can you give me an example of sftp\_open''), and responses or feedback to wizard turns (e.g., ``Yes.''). We analyze the distribution of these different behaviors in greater detail in \Cref{sec:annotation_results}.

We noted that a number of programmers endeavored to access more traditional API resources to complete the tasks, e.g., by requesting that Apiza direct them to the official documentation. Generally, these requests came midway through the study, indicating that the programmers may have gotten stuck or become frustrated with the system, and either believed they would be unable to procure the necessary information via directed questions or decided they would prefer to fall back on  standard approaches to API navigation. One programmer (in Dialogue \#2) began the session by immediately querying ``libssh documentary [sic]''; later, the programmer requested that Apiza ``list all function of libssh documentary''. In the meantime, the programmer also asked more targeted questions about specific functionality and components. 

Finally, while statistically characterizing the effects of programmer experience on their behavior and performance is outside of the scope of this study (see \Cref{sec:prob}), we will remark that programmers who reported \textgreater 10 years of programming experience completed more tasks, on average, than those with less experience, and those who reported ``Intermediate'' familiarity with the API domain in the libssh scenario\footnote{We only mention libssh here, as only 2 programmers reported ``Intermediate'' or higher familiarity in the Allegro scenarios} completed more tasks than those with ``Novice''-level familiarity (these figures can be found in the \texttt{experimental\_results/} folder in our online Appendix).

\subsubsection{Programmer expectations} 

In the dialogues, we observed habits and expectations that programmers may have developed over time by interacting with other virtual assistants and search tools. Programmers often avoided the use of pronouns to refer to recently mentioned API components. They may have wanted to be as clear as possible in the event that Apiza would be unable to resolve the antecedent. Some programmers treated Apiza like a more traditional search engine by constructing queries as ungrammatical sequences of keywords (e.g., queries like ``verify ssh server example code'' and ``Allegro play sound'').

At the same time, many programmer questions demonstrated greater expectations for Apiza to be able to assess programming scenarios and recall previous conversational context than would be expected of a traditional search engine. For instance, the programmer in Dialogue \#24 asked ``if I have an ALLEGRO\_EVENT called ev, should ev.type == ALLEGRO\_EVENT\_KEY\_DOWN return true when I press a key?''. This question demonstrates the programmer's expectation that Apiza would be able to not only comprehend variable assignment, but also to evaluate a high level design pattern using that custom variable. 

Similarly, the programmer in Dialogue \#6 posited a question about a code snippet in a previous turn:

\begin{excerpt}
\label{exc:followup}
\begin{tcolorbox}[left=5pt,right=-15pt,top=5pt,bottom=5pt]
\begin{flushleft}
\begin{dialogue}
\speak{Wiz} ... Example: ...
while ((rc = channel\_read(channel, buffer, sizeof(buffer), 0)) \textgreater 0) \{ ...
\vspace{0.1cm}
\speak{Pro} what is buffer in that code sample?
\vspace{0.1cm}
\speak{Pro} just a char *?
\end{dialogue}
\end{flushleft}
\end{tcolorbox}
\label{exc:followup}
\end{excerpt}

The programmer did not directly ask Apiza to give the type of a particular parameter in a particular function. Rather, the programmer expected Apiza to recall the code snippet sent in the previous message, identify that ``buffer'' refers to a variable name used in that snippet, and determine the type of that variable based on its usage as a parameter in certain function invocations.

Finally, we observed that programmers occasionally asked questions outside of the scope of the system's functionality as described in the scenario descriptions. Some of these questions, such as ``How's it going?'' and ``How is the weather where you are?'' related to social functions that did not serve to communicate task-relevant information. It is not necessarily clear that programmers \emph{expected} the system to respond to these types of questions, as these questions may have been performative, with programmers aware that their dialogues would eventually be reviewed, or experimentative, with programmers attempting to explore the range of the system's functionality.

In other cases, programmers chose to direct general C syntax questions to Apiza (such as ``how to print to stdout ?'' in Dialogue \#3), despite being allowed to search the internet for this information. These questions would seem to demonstrate an expectation that the system be able to comprehend more general programming concepts unrelated to the target API, or at least be able to identify the programmers' intent and refer them to useful external resources.

\subsubsection{Wizard strategies}
We did not observe that different Wizards clearly performed ``better'' or ``worse'' than other wizards. As with programmers, detailed comparisons of the performance of different wizards' dialogue strategies are not in the scope of the study (see \Cref{sec:prob}). However, we do make qualitative observations regarding the diversity of strategies used.

When faced with API usage questions, wizards generated a range of response types. For instance, in response to common ``How?'' questions eliciting guidance to implement certain objectives, wizards sometimes gave detailed pattern descriptions and examples, suggestions to utilize certain API functions (e.g., ``al\_get\_mouse\_event\_source will give you the mouse event source''), lists of relevant functions, requests for clarification (e.g., ``There are several authentication functions. How would you like to authenticate?''), or statements that the request could not be fulfilled (e.g., ``I am sorry, I am confused.''). \Cref{sec:annotation_results} provides a more comprehensive overview of the response types observed. 

Notably, even for questions that were nearly identical, wizards produced different types of responses in different dialogues. For instance, the programmers in Dialogues \#21, \#26, and \#28  all asked similar forms of the question ``How do I get keyboard input?'' as their first API-related question in the dialogue. However, the wizards all produced different dialogue act types in response: the wizard in Dialogue \#21 provided an excerpt of documentation relating to a static variable that mentioned ``keyboard'' and ``pressed'' ; the wizard in Dialogue \#26 indicated that several functions would be required, and provided the first one; and the wizard in Dialogue \#28 gave a thorough code example. 

Because the wizards were responding to the first API-related question in all three dialogues, they were all making decisions in the context of the same essential ``dialogue state.'' Yet, each responded with a different dialogue act type, which demonstrates that they were using different dialogue strategies. Furthermore, we note that the same wizard participated in both Dialogue \#21, and Dialogue \#26. The fact that the wizard chose to use a different dialogue strategy in Dialogue \#26 that immediately provided a relevant example, rather than the documentation for a component with related keywords, is indicative of the learning effect that we expected to produce.

\subsubsection{Deviations in dialogue structure}
As noted in \Cref{sec:dialogues}, the interactions in the dialogues typically followed a simple question-answer structure, in which the programmer sought information and the wizard provided it. This was not always the case, however. Sometimes wizards sought information from the programmer (e.g., clarification questions); other times, programmers proactively offered information. Occasionally, a speaker would send multiple messages in a row, as demonstrated in the following excerpt from Dialogue \#24:

\vspace{-.2cm}
\begin{excerpt}
\label{exc:followup}
\begin{tcolorbox}[left=5pt,right=-15pt,top=5pt,bottom=5pt]
\begin{flushleft}
\begin{dialogue}
\speak{Pro} if I have an ALLEGRO\_EVENT called ev, should ev.type == ALLEGRO\_EVENT\_KEY\_DOWN return true when I press a key?
\vspace{0.1cm}
\speak{Wiz} I’m not sure I understand
\vspace{0.1cm}
\speak{Wiz} Are you interested in al\_get\_keyboard\_state or al\_key\_down?
\vspace{0.1cm}
\speak{Pro} no
\vspace{0.1cm}
\speak{Pro} tell me about al\_get\_keyboard\_state
\end{dialogue}
\end{flushleft}
\end{tcolorbox}
\label{exc:followup}
\end{excerpt}

In total, $15.7\%$ of the wizards' messages and $18.3\%$ of the programmers' messages were followed by one or more messages from that speaker. Furthermore, speakers often expressed multiple, separate ideas and intentions within a single message. Identifying these utterances, parsing their meanings, and determining appropriate response types are key challenges a virtual assistant needs to overcome. \Cref{sec:annotation} describes the first steps we've taken towards tackling these challenges. 

\subsection{Threats to Validity}

As in any experimental study, our experimental design carries a number of threats to validity, including human factors, the participant selection process, the details of the experimental scenarios, participants' adherence to guidelines, and the design of the communication interface.  Here, we will briefly address these threats, and explain the steps we took to mitigate them.

\smallskip

\textbf{Human factors.} Human factors that were not controlled in the experimental design had the potential to impact participants' behavior and performance. Factors such as fatigue and distraction were not possible to observe remotely; therefore, efforts were made to mitigate them. To reduce the effects of fatigue and distraction, we limited each experimental session to 90 minutes. We also attempted to curb distraction by providing distraction-free virtual environments to both the programmers (the virtual machine testing environment) and the wizards (the custom tool). Still, those factors may have played a role in some experimental sessions.

\smallskip

\textbf{Participant selection.}  We invited programmers from a wide range of backgrounds to participate, in order to observe a wide range of programmer behavior. However, this leniency means that it is possible that the behavior observed may have been different with a different set of programmers. Similarly, while our decision to recruit a fairly large number of wizards (6) and allow them to participate for a variable number of studies was intended to promote diverse wizard strategies, it introduced more factors that may have affected the results of each experimental session. We attempted to mitigate these risks by recruiting a fairly large number of programmers (30) and having all participants complete entry surveys to document their backgrounds. Still, there were a large number of individual differences among participants that were not balanced in our experimental design, meaning that certain statistical analyses of the corpus may be confounded by these factors. That said, we believe that our participant selection process ultimately struck a reasonable balance between experimental control, generalizability, and practicality.

\smallskip

\textbf{Experimental Scenarios.} The choice of APIs and the design of the particular tasks we asked participants to complete in each scenario may have also had an effect on the types and quantities of interactions that occurred. The domain of each API and the quality of its documentation likely affected the wizards' ability to find relevant components and make recommendations, as well as the programmers' ability to determine which components were appropriate and how to use them. Similarly, the specificity and wording of the tasks may have influenced the types of questions programmers asked, or primed them with certain keywords. However, there is no single API or task set can generalize to all APIs and tasks. Therefore we chose two APIs from different domains and made one task set more open-ended than the other. The tasks themselves were designed to cover a relatively broad range of the functionality offered by the APIs. We kept the task descriptions short, and avoided identifying specific API components or implementations when possible. Still, the many differences between the two scenarios (such as the different number and quality of tasks) means that any differences observed between them in the results cannot be attributed to any individual variable.

\smallskip

\textbf{Adherence to Scenario Guidelines.}
To encourage programmers to follow the scenario guidelines, we included all pertinent information and restrictions in the document detailing the experimental tasks, and before beginning sessions, we asked them to read through the document and gave them the opportunity to ask any questions. Nevertheless, some programmers failed to adhere to guidelines relating to the task completion order in the libssh scenario and the prohibition on online API resources. As shown in \Cref{tab:taskcompletion}, several programmers moved on to subsequent tasks in the libssh scenario without successfully completing the previous ones as instructed in the task document. Additionally, 5 programmers wound up making API-related internet queries; 2 of them made searches and accessed resources related to a single API component, 2 of them accessed a tutorial, and 1 asked questions about numerous components and accessed a tutorial. 

However, we do not believe these instances invalidate the results of those experimental sessions with respect to the downstream dialogue strategy development task. The various restrictions were not intended as experimental controls; they were primarily intended to promote programmer-wizard interaction. By violating the restrictions, programmers may or may not have wound up interacting with Apiza less than they otherwise would have, but the interactions that they did have are still relevant towards dialogue strategy development. Therefore, we chose to keep these sessions in the corpus and annotate them as normal. That said, future attempts at statistical analysis should be aware of these cases, and may consider excluding them altogether.

\textbf{Communication interface.} Finally, the chosen communication interface may have impacted the results of the study. Being limited to text-based communication over Slack may have affected the amount of information wizards decided to include in individual messages (e.g., they may have suggested fewer API components in spoken messages, or more components if they had been able to link directly to documentation). At the same time, the interface may have affected the frequency and wording of programmer questions. As analyzing the differences between communication modalities was not in the scope of this study, we felt that text-based communication via Slack would suffice, as Slack is a fairly well-known messaging platform with few restrictions on message length or frequency, and one for which many developers have created actual bots.

\section{Corpus Annotation}
\label{sec:annotation}

We annotated the dialogue acts in the Wizard of Oz corpus along four highly-generalizable dimensions. As described in ~\Cref{sec:strats}, dialogue act annotations are needed to train virtual assistants to perform specific tasks. Different annotation schemes are needed to facilitate different functionality in a virtual assistant. Future researchers may find that they need to annotate additional or alternative dimensions to capture certain types of interactions or functionality observed in the corpus. To attempt to generate a comprehensive list of all dimensions for all functionality is not in line with findings from related literature~\cite{kang2018data}.  Instead, we provide a first round of highly-generalizable annotations as a foundation on which future virtual assistant developers can build (as described in \Cref{sec:das}).

\subsection{Research Questions}

We investigated four specific research questions:

\smallskip
\begin{description}
\item[$RQ_1$] What is the composition of the corpus in terms of \textbf{illocutionary} dialogue act types?
\end{description}
\smallskip

An \emph{illocutionary dialogue act type} describes the illocutionary function of an utterance (e.g., distinguishing between a statement intend to inform and a statement intended to suggest a future action). The rationale for $RQ_1$ is to discover the conversational ``flow'' of the API dialogues. Annotating illocutionary dialogue act types enables us to train a virtual assistant to identify the intent behind a user's utterance and to predict the appropriate type of response (e.g., whether to respond to a question with an answer or a follow-up question).  

\smallskip
\begin{description}
\item[$RQ_2$] What is the composition of the corpus in terms of \textbf{API} dialogue act types?
\end{description}
\smallskip

Whereas illocutionary dialogue act types label utterances as ``questions'' or ``statements,'' \emph{API dialogue act types} describe what types of API knowledge (such as functionality, usage patterns, or examples) are addressed in each utterance. The rationale for $RQ_2$ is to evaluate the domain-specific content of the API dialogues.  As Wood~\emph{et al.}~\cite{wood2018detecting} point out, a virtual assistant must be able to identify the type of domain-specific knowledge a programmer asks about in order to respond with a relevant answer.

\smallskip
\begin{description}

\item[$RQ_3$] What is the composition of the corpus in terms of \textbf{backward-facing} dialogue act types?
\end{description}

\smallskip

A \emph{backward-facing dialogue act type} describes how an utterance relates to a previous utterance (e.g., supporting or rejecting a prior utterance). The rationale for $RQ_3$ is to observe relationships among the utterances in the API dialogues. We found that programmers and wizards did not always engage in simple question-answer turn-taking. Rather, dialogues sometimes became complex, with participants referencing prior turns and multiple conversational ``threads'' progressing in parallel. A virtual assistant should be able to track dialogue threads in order to draft contextually appropriate responses.

\smallskip
\begin{description}
\item[$RQ_4$] What is the \textbf{traceability} of specific API components in the corpus?
\end{description}
\smallskip

The rationale behind $RQ_4$ is to identify and track the specific API components that are addressed throughout the dialogues. This identification is related to the concept assignment problem, as it involves connecting specific software components to their relevant natural-language concepts. We refer to this identification as \emph{traceability}, as it relates to the concept of traceability in software engineering~\cite{1041053}.  A virtual assistant for APIs must be able to identify the API elements relevant to a user's utterance (even when those elements are not mentioned by name).

\smallskip

We addressed each of these research questions by annotating the conversation corpus along a different dimension of an annotation scheme.

\vspace{-0.2cm}
\subsection{Methodology}
\label{sec:methodology}

This section details the annotation scheme used to investigate the research questions presented above. The complete lists of dialogue act types for all annotation dimensions are given in Appendix B.

\subsubsection{Segmentation}
\label{sec:segmentation}
Before assigning any labels, we first had to segment the Slack messages into individual segments, or \emph{utterances}.  McCowan~\emph{et al.}~\cite{mccowan2005ami} emphasize that dialogue acts reflect speaker intention, and in their guide, they recommend that ``each time a new intention is expressed, you should mark a new segment.'' The length of an utterance is variable; it may consist of a single word or entire paragraphs, depending on the speaker's intention. In our experiments, programmers and wizards often expressed multiple intentions, responded to multiple utterances, or referenced multiple API components in the course of a single message. Therefore, it was important to segment the messages before labeling. 

Our corpus presented a unique segmenting challenge: wizards often shared verbatim chunks of documentation to the programmer. These chunks could contain several paragraphs worth of information. To segment each individual utterance of the verbatim documentation would not quite be appropriate, as they do not necessarily represent separate intentions on the part of the wizard. On the other hand, to group an entire chunk of documentation into a single segment would be too coarse grained, as wizards had the ability to purposefully include and exclude certain types of information using the documentation tool described in \Cref{sec:wiz}. Given that fact, we chose to segment chunks of documentation by topic (e.g., discussion of all parameters was segmented into an utterance, separate from discussion of the return values).

\subsubsection{Methodology of RQ1: Illocutionary Annotation}
We labeled illocutionary dialogue act types using the so-called ``AMI labels'' (see Table 7 in Appendix B). The AMI corpus, presented by McCowan \emph{et al.}~\cite{mccowan2005ami}, provides a coarse-grained set of labels for illocutionary dialogue act types that are applicable to many types of conversations.  As Gangadharaiah~\emph{et al.}~\cite{gangadharaiah2018we} point out, a useful place to start annotation is with a set of 10-20 coarse-grain labels to provide a common comparison point with other datasets, even though these annotations alone are not sufficient for industrial use.

Our methodology for annotation was straightforward: McCowan \emph{et al.}~\cite{mccowan2005ami} provide an annotation guide with detailed instructions for every dialogue act type.  We followed this guide for every conversation, labeling each utterance with the most appropriate label.  We exclude three of the original AMI labels (STALL, FRAGMENT, and BACKCHANNEL) that were primarily relevant to spoken modalities. Note that we annotated both sides of the conversations, wizard and programmer, even though a virtual assistant would only need to classify the dialogue acts of the programmer -- it would know the wizard's dialogue act types because it would have generated them.  However, we annotate both sides anyway, since we are interested in the wizards' conversation strategies and providing guidance to designers of virtual assistants.  This decision is in line with recommendations by Wood~\emph{et al.} in a study of software engineering virtual assistants~\cite{wood2018detecting}.

We compared the frequencies of illocutionary dialogue act types in our corpus to one corpus in a similar domain, the Wizard of Oz debugging corpus~\cite{wood2018detecting}, and one corpus in a dissimilar domain, the AMI meeting corpus~\cite{mccowan2005ami}. The debugging corpus consists of written dialogues between programmers and wizards completing a series of bugfixing tasks, while the AMI corpus consists of spoken multi-party discourse in simulated meetings. 

These comparisons are valuable because the lack of domain-specific training data has been a major barrier in the development of virtual assistant technology for software engineering tasks. By highlighting similarities across domains, we hope to encourage researchers to consider ways to leverage existing data; for instance, if the distribution of illocutionary dialogue act types in API usage dialogues is evidently similar to that in similar domains, it may encourage researchers to consider using certain transfer learning techniques for tasks such as dialogue act classification. Conversely, stark differences between the corpora may warn researchers against incorporating out-of-domain data into certain models. Of course, the two corpora chosen for comparison are by no means representative of the vast collection of publicly-available conversational data; however, they allow us to characterize our corpus in the context of similar and dissimilar domains.

\subsubsection{Methodology of RQ2: API Knowledge Annotation}
The API dialogue act types in each utterance were labeled according to the taxonomy proposed by Maalej and Robillard~\cite{maalej2013patterns} (see Table 8 in Appendix B). In their work, they generated a set of 12 broad categories that may be used to classify the information types present in API documentation. In a different paper, Tian~\emph{et al.}~\cite{tian2017apibot} applied these labels not only to API documentation, but to questions about documentation as well. By training a model on API question/answer pairs associated with these labels, they were able to achieve high performance on an API information type retrieval task.

We followed the annotation guide provided by Maalej and Robillard, which describes in detail every knowledge type and provides suggestions to resolve uncertainties. We labeled API dialogue act types in utterances by both the wizards and the programmers, but we only labeled utterances that actually contained API information. Unlike the illocutionary dialogue act annotation scheme, Maalej and Robillard explicitly allowed for multiple labels to be applied to a single unit; therefore, when appropriate, we annotated utterances with multiple API dialogue act types. 

We found that the knowledge type that Maalej and Robillard referred to as ``NON-INFORMATION'' actually encompassed valuable information in the context of these dialogues, such as the names of functions, their parameters, and their return values. As such, we decided to create a ``BASIC'' category for utterances meant to convey these basic pieces of information.

\subsubsection{Methodology of RQ3: Backward-Facing Annotation}
We used another layer from the AMI annotation scheme~\cite{mccowan2005ami} to capture the relationships among utterances (see Table 9 in Appendix B). This layer consists of only four backward-facing dialogue act types: ``POSITIVE,'' ``NEGATIVE,'' ``PARTIAL,'' and ``UNCERTAIN.'' Each of these covers a wide range of relationship types; for instance, ``POSITIVE'' can imply agreement with a previous utterance, understanding of a previous utterance, or an attempt to provide something that a previous utterance requested.

We again followed the AMI guidelines. In this scheme, an utterance can only relate directly to a single prior utterance, so an annotation consisted of a single label and the ID of the related utterance. Any utterance could relate to any prior utterance by either speaker. Any utterance without a backward-facing function (e.g., an utterance starting new lines of questioning) was given a placeholder ``[NONE]'' label.

In the course of annotating relationships, we found that the four backward-facing dialogue act labels did not provide enough granularity to satisfactorily capture all of the observed relationships between utterances. We noted that other dialogue act labeling schemes (such as DAMSL~\cite{Core1997CodingDW}) provide more distinction between different types of relationships (e.g., DAMSL distinguishes among ``Accept,'' ``Acknowledge,'' and ``Answer'' relationships, all of which AMI would classify as ``POSITIVE''). Therefore, we chose to add three additional labels to the original AMI set of backward-facing dialogue act type labels in order to more meaningfully evaluate the relationships that occur in our specific corpus.

The three labels that we added were ``REPEAT,'' ``FOLLOW-UP,'' and ``CONTINUE.'' The ``REPEAT'' label was used to mark repetitions, repairs, or rephrasings of a previous utterance. The ``FOLLOW-UP'' label was used to mark acts that followed up on a previous utterance, either by asking a question predicated on the utterance, or providing unprompted information or suggestions based upon the previous utterance. The ``CONTINUE'' label was intended to link contiguous utterances by one speaker that form a single question or response. For instance, Apiza often responded to requests for documentation by sending information about parameters, return values, functionality, etc. These responses each comprised several utterances -- it would be inaccurate to label each utterance as having a totally separate relationship to the original query.

\subsubsection{Methodology of RQ4: Traceability Annotation}
The ``traceability'' annotation of specific API components was inspired by topic- and entity-labeling methodologies in NLP and SE~\cite{stoyanov2008topic,wiebe2005creating, witte2008text}. This dimension did not involve dialogue act types per se -- instead, it used the names of API components as labels. We went through every utterance and labeled any API components that were the topic of the utterance. An API component did not need to be referenced by name to constitute a topic, and direct references to a component did not necessarily make them a topic. Rather, the decision was heavily based on context, the intention of the speaker, and the retroactive role of the component in the conversation. 

For instance, the utterance ``How do I ensure a session was successfully created?'' does not explicitly name any component of the libssh API -- however, the ``session'' is in reference to an \texttt{ssh\_session} struct, and the question is asked directly after a discussion of the \texttt{ssh\_new()} method. Therefore, both the \texttt{ssh\_session} struct and the \texttt{ssh\_new()} method are labeled for that utterance.

We referred to the official documentation for each API to determine which API components to include as potential labels. For the Allegro API, we included all API components with indexed pages in the documentation\footnote{https://liballeg.org/a5docs/5.2.2/index\_all.html} -- 338 in total. For the LibSSH API, we included the functions from several modules as well as all data structures listed in the documentation\footnote{https://api.libssh.org/master/modules.html} -- 1038 in total. We include these complete lists in our online Appendix (see ~\Cref{sec:conclusion}).

\begin{figure*}[t]
  \includegraphics[width=\linewidth]{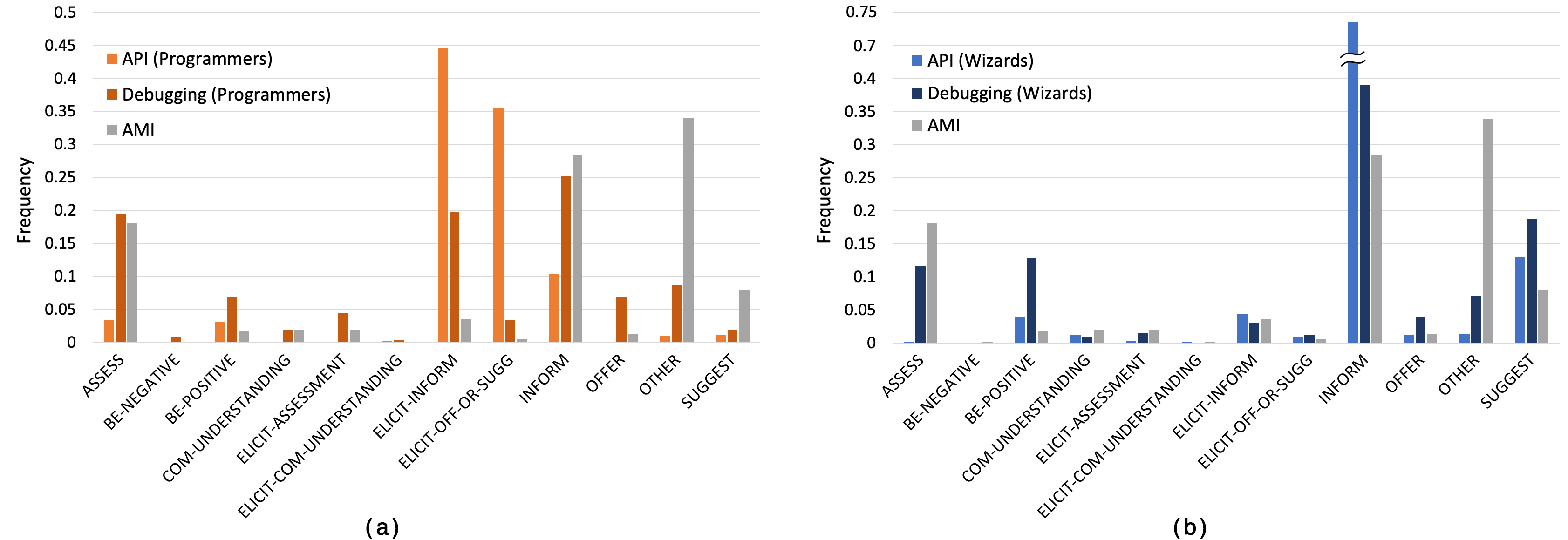}
   \caption{Frequencies of illocutionary dialogue act types in our API corpus, the debugging corpus by Wood~\emph{et al.}~\cite{wood2018detecting}, and the AMI meeting corpus\cite{mccowan2005ami}. (a) shows frequencies across programmer utterances, while (b) shows frequencies across wizard utterances.}
  \label{fig:DAdist}
\end{figure*} 
\subsubsection{Note on Reliability}	

In any annotation process, it is important to consider the \emph{reliability} of the annotations, or ``the extent to which different methods, research results, or people arrive at the same interpretations or facts''~\cite{krippendorff2011agreement}. Bias, fatigue, and other factors may cause an individual annotator to produce inconsistent or unreliable results~\cite{hsieh2005three,downe1992content}. 

It is common for researchers to gauge the reliability of their annotations by asking multiple, independent annotators to annotate data and then calculating an agreement score using, e.g., Cohen's kappa or Krippendorff's alpha~\cite{krippendorff2004reliability}. Establishing reliability is especially important in ``conventional'' qualitative analysis, or ``open-coding'' processes in which annotators do not use predetermined sets of labels. In those cases, agreement among independent annotators does not just indicate the reliability of a particular set of annotations, but rather, the reliability of the annotation scheme as a whole.

However, the act of calculating agreement does not itself improve reliability. Furthermore, agreement scores are notoriously difficult to interpret (e.g., while an agreement score of $.8$ is generally considered to indicate high reliability, it is not sufficient for applications that are ``unwilling to rely on imperfect data''\cite{craggs2005evaluating}). 

By contrast, we performed ``directed'' qualitative analysis; that is, we annotated the corpus using preexisting sets of labels. Our priority was not to measure the reliability of the existing annotation schemes, but to ensure the correct application of those schemes. To achieve unbiased results in this sort of analysis, Hsieh and Shannon~\cite{hsieh2005three} suggest using an ``auditing'' process, in which experts discuss the application of label sets and resolve any ambiguities. This type of procedure has frequently been used in the social sciences~\cite{BENGTSSON20168,spooren2010coding}, and more recently, in software engineering~\cite{wood2018detecting}. This process allows for the creation of a single, higher-quality set of annotations for use in applications that are less willing to rely on imperfect data~\cite{wood2018detecting}. 

We followed this procedure: the first author annotated the corpus following the guidelines for existing annotation schemes for each relevant dimension. Whenever there was some ambiguity as to the correct application of an annotation scheme, the first and third authors discussed the situation and decided on a correct implementation (or, in some cases, modifications to the scheme). Although this auditing process does not allow us to calculate an agreement score, it allowed us to generate a single set of ``more accurate'' annotations~\cite{hsieh2005three}. Still, we acknowledge that future researchers who wish to apply these annotations directly may be wary of the fact that a reliability metric cannot be calculated.

\section{Annotation Results}
\label{sec:annotation_results}

Next, we discuss the results of the annotation process described in the previous section. Before assigning any labels, we segmented the programmers' and the wizards' messages into discrete utterances, as described in \Cref{sec:segmentation}. We ultimately segmented the 1947 messages in the corpus into 3183 utterances. Programmers' messages contained 1.1 utterances on average, with 6\% containing more than one utterance. Wizards' messages contained 2.2 utterances on average, with 57\% containing more than one utterance. 

\subsection{RQ1: Illocutionary Dialogue Act Types}

Interactions in the API dialogues most frequently consisted of questions levied by the programmers, and information or suggestions provided by the wizards. Other illocutionary types and patterns emerged, but they were infrequent in this corpus compared to the more conversational AMI and debugging corpora.

\Cref{fig:DAdist} shows the composition of the corpus in terms of illocutionary dialogue act labels. Programmers most frequently used dialogue acts of the ELICIT-INFORM and ELICIT-OFFER-OR-SUGGESTION illocutionary types, collectively accounting for approximately $80\%$ of all programmers' dialogue acts.  Wizards primarily used dialogue acts of the INFORM type, accounting for nearly $74\%$ of their dialogue acts. The next most common label for the wizards was SUGGEST, accounting for $13\%$ of their labels.  These preferences seem to reflect the task goals and the participants' different roles in the API dialogues; programmers sought to learn how to use the API, and wizards provided the desired information.

We compared the distribution of AMI labels in our corpus to two others: the AMI meeting corpus~\cite{mccowan2005ami} and the Wizard of Oz debugging corpus~\cite{wood2018detecting}.  As shown in \Cref{fig:DAdist}, the three corpora shared a few traits: the relatively high frequency of INFORM acts and the relative rarity of the BE-NEGATIVE, COMMENT-ABOUT-UNDERSTANDING, and ELICIT-COMMENT-ABOUT-UNDERSTANDING acts. 

Beyond those similarities, the distributions varied substantially among the three corpora. Compared to the speakers in the AMI corpus, programmers in both Wizard of Oz studies used more dialogue acts of the ELICIT-INFORM and ELICIT-OFFER-OR-SUGGESTION types and fewer of the INFORM and SUGGEST types, while the opposite was true of the wizards. These tendencies were much more pronounced in the API study than the debugging study; wizards in the API study used INFORM acts nearly twice as frequently as wizards in the debugging study, and programmers used ELICIT-INFORM acts more than twice as frequently. Other notable differences include the relative lack of ASSESS and OFFER act types in the API corpus compared to the other corpora.

The distribution of illocutionary dialogue act types in the API corpus is highly imbalanced, even when compared to that of the other Wizard of Oz corpus in the software engineering domain. This imbalance has a few implications for researchers looking to build an intelligent virtual assistant for API usage. On one hand, it is more difficult to train robust dialogue act models on skewed datasets~\cite{doi:10.1162/089120100561737}. For instance, a classifier trained on this corpus would be unlikely to correctly identify an utterance of the OFFER type, as it would have been exposed to very few training instances of that type. Researchers have previously used techniques such as downsampling~\cite{doi:10.1162/089120100561737}, oversampling, and SMOTE\cite{wood2018detecting} to counteract the effects of imbalanced data in dialogue act modeling, but these solutions are typically poor replacements for real data. In terms of dialogue strategy training, this imbalance may reduce the fidelity of a simulated learning environment, ultimately leaving a virtual assistant under-equipped to handle certain dialogue states.

On the other hand, this imbalance means that virtual assistant designers can narrow the range of interaction types supported by a virtual assistant for API usage while still providing the desired functionality. Because a few key illocutionary dialogue act types account for the vast majority of turns, virtual assistant designers may choose to focus on those types and group the rest into the OTHER category. Doing so would inevitably reduce the expressiveness of the dialogue system, but it would ultimately be easier to train due to the constrained state and action space~\cite{rieser2011reinforcement}.

\subsection{RQ2: API Dialogue Act Types}
\label{sec:api_inf}

\begin{figure}[t]
  \includegraphics[width=\linewidth]{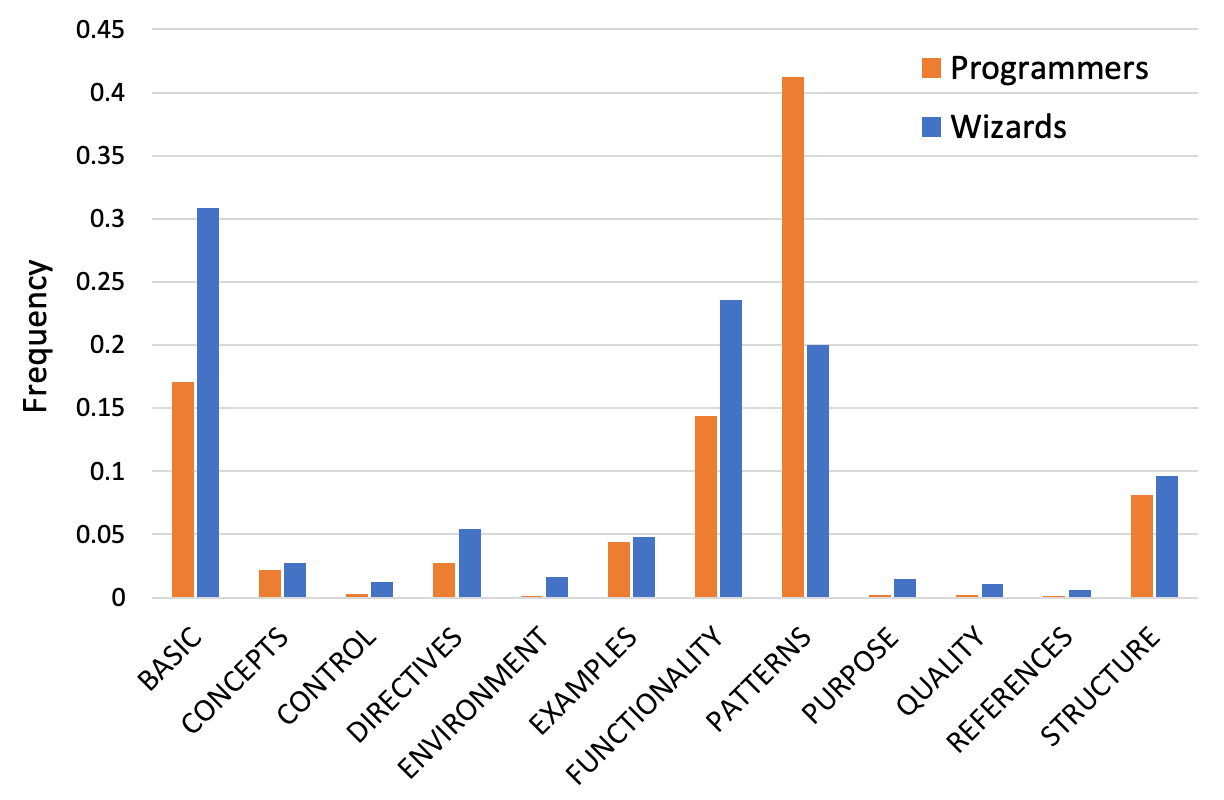}
  \vspace{-.8cm}

   \caption{Frequencies of API dialogue act types in the corpus.}

  \label{fig:APIdist}
\vspace{.6cm}
\end{figure}

\begin{figure}[t]
  \includegraphics[width=\linewidth]{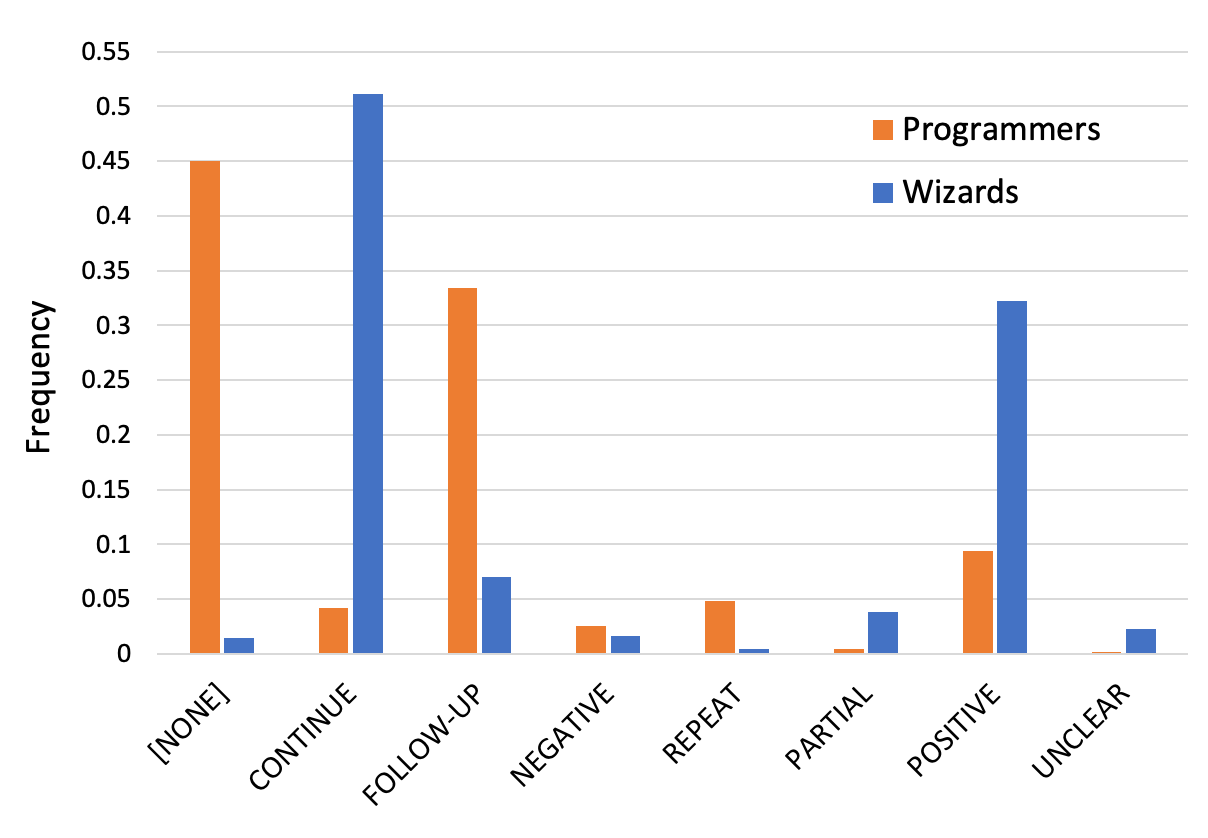}
  \vspace{-.8cm}
   \caption{Frequencies of backward-facing dialogue act types in the corpus.}

  \label{fig:BFdiff}
\end{figure}

The overwhelming majority of interactions related directly to the API usage task. We annotated 2668 utterances with API dialogue act types across 1741 messages; 89\% of programmers' messages and 90\% of wizards' messages were labeled with at least one API dialogue act type. \Cref{fig:APIdist} shows the composition of the corpus in terms of API dialogue act labels.

The dialogues focused heavily on three API information types: PATTERNS, FUNCTIONALITY, and BASIC. Programmers most frequently asked for information about PATTERNS (accounting for over 40\% of their utterances), describing \textit{how} to accomplish a specific objective with the API, followed by queries about BASIC information and FUNCTIONALITY details.  Wizards referenced the same three API types most frequently, but the distribution differed: BASIC information was present in about 30\% of the wizards' utterances, FUNCTIONALITY in about 25\%, and PATTERNS in about 20\%. 

Wizards and programmers both used all twelve API dialogue act types. However, five types occurred three or fewer times across all programmer utterances: PURPOSE, REFERENCES, CONTROL, QUALITY, and ENVIRONMENT.  These were also the least-frequently-occurring types in wizard utterances.

Again, we observe an imbalanced distribution of API dialogue act types. This imbalance presents researchers with challenges and opportunities similar to those discussed in RQ1. Specifically, while it would be difficult to train some ML models on this skewed dataset, virtual assistant designers may choose to provide functionality for only a subset of the most common API dialogue act types. We discuss these implications in greater detail in \Cref{sec:discussion}.

\subsection{RQ3: Backward-Facing Dialogue Act Types}
\label{sec:relationships}

Programmer-wizard interactions frequently took place in the context of previous utterances.

\Cref{fig:BFdiff} shows the distribution of backwards-facing dialogue act types. The majority of the programmers' dialogue acts were assigned either the [NONE] label (meaning they did not have a direct relationship to any previous utterance) or the FOLLOW-UP label. The PARTIAL, NEGATIVE, and UNCLEAR tags saw minimal use by the programmers. The majority of the wizards' dialogue acts were assigned the CONTINUE label, primarily due to the segmentation method used for verbatim chunks of documentation. The next most frequent label was POSITIVE, and the least frequent labels were [NONE] and REPEAT. 

Wizards often relied on previous conversational context to produce an appropriate response. The following excerpt demonstrates one such situation:

\begin{excerpt}
\begin{tcolorbox}[left=5pt,right=-18pt,top=5pt,bottom=5pt]
\begin{flushleft}
\begin{dialogue}
\speak{PRO} Apiza, what is the command to create a new ssh\_session in the libssh API?
\vspace{0.1cm}
\speak{WIZ} The command to create a new ssh session is: ssh\_session ssh\_new(void) [...] Returns A new ssh\_session pointer, NULL on error.
\vspace{0.1cm}
\speak{PRO} Just to confirm, is the ssh\_session type a pointer type?
\vspace{0.1cm}
\speak{WIZ} ssh\_session is not a pointer type.
\vspace{0.1cm}
\speak{PRO} But when you gave me the command to create a new ssh session, you said it ``Returns A new ssh\_session pointer...''. I'm confused.
\vspace{0.1cm}
\speak{WIZ} ssh\_new returns a pointer to type ssh\_session
\end{dialogue}
\end{flushleft}
\end{tcolorbox}
\label{exc:bouncy}
\end{excerpt}

In that exchange, the wizard needed to recall the context of the method \texttt{ssh\_new} as well as the question ``is the ssh\_session type a pointer type'' in order to rectify the programmer's confusion. Furthermore, programmers did not necessarily respond to the wizard's most recent message; 16\% of programmers' backward-facing dialogue acts were related older to messages in the conversation.

\subsection{RQ4: Traceability}

In total, we identified 145 specific API components discussed across 2221 utterances. The total number of API component labels and number of unique component labels annotated are shown in \Cref{tab:trace}.

The wizards referred to specific API components more often than the programmers. This trend can be viewed in the context of the dialogue pattern observed in \Cref{sec:dialogues} and the dialogue act type distributions for the other dimensions; specifically, programmers often elicited information about known components or used descriptions of desired API functionality to elicit recommendations, and wizards responded with information about specific components. We observe that a minority of programmer's utterances labeled with the ELICIT-OFFER-OR-SUGGESTION illocutionary dialogue act type (39\%) are labeled with a traceability label, while the majority of wizard utterances labeled with the SUGGEST type (85\%) do contain a traceability label. Meanwhile, a majority of programmer utterances labeled with the ELICIT-INFORM illocutionary type (70\%) have at least one traceability label (i.e. target a specific API component), and most wizard utterances labeled with the INFORM type (87\%) have traceability labels.

We also observe that more traceability labels were applied to utterances in the libssh scenario (particularly in programmer utterances), while a broader range of distinct labels were applied to utterances in the Allegro scenario. There are many factors that may have affected these trends, including the design, domain, and size of the different APIs, the tasks given in the different scenarios, and individual differences among wizards and programmers. 

\begin{table}[t]
	\centering
  \caption{Number of references to specific API components in the corpus. 338 components in the libssh API and 1038 in the Allegro API were available as traceability labels.}
  \label{tab:trace}
	\begin{tabular}{llccc}
		\toprule
		 Library & Variable & Programmer & Wizard & Both \\ 
		\midrule
        \multirow{2}{*}{libssh} & \% Messages w/ labels & 60.4     & 79.6    & 69.6  \\
		& \# Labels & 358 & 1007      & 1365  \\
		& \# Distinct labels & 37     & 59    & 59  \\
		\midrule
        \multirow{2}{*}{Allegro}& \% Messages w/ labels & 47.0     & 77.1    & 62.2  \\
		& \# Labels & 240 & 801      & 1041  \\
		& \# Distinct labels & 54     & 86    & 86  \\
		\midrule
        \multirow{2}{*}{Both}& \% Messages w/ labels & 54.3     & 78.4    & 66.1  \\
		& \# Labels & 598 & 1808      & 2406  \\
		& \# Distinct labels & 91     & 145    & 145  \\
		\bottomrule
	\end{tabular}
\end{table}

\section{Discussion}
\label{sec:discussion}

This section summarizes several key insights from our study and provides specific recommendations for future researchers and virtual assistant developers.

\subsection{Insights}
\label{sec:insights}
We synthesize the following insights from observations of the Wizard of Oz experiments and dialogue act annotations in the context of the broader research landscape.

\subsubsection{Value of the Wizard of Oz Methodology}

We support the use of the Wizard of Oz methodology to collect dialogues in which programmers and wizards exhibit a broad range of realistic behaviors and strategies. In these experiments, the dialogues were dense with API-related interactions; over 90\% of programmers' messages were directly related to some aspect of the API. The rest were either related to dialogue control (e.g., salutations, stalling, expressing gratitude), served other task-related ends (e.g., questions related to C syntax) or intended to probe the system's capabilities (e.g., ``It was hot here today. How is the weather where you are?''). As we noted in \Cref{sec:obs}, even these interactions are valuable to observe, as they are expositive of programmers' expectations for the target system, and should be considered when deciding on the scope of a virtual assistant's functionality.

We note that most Wizard of Oz experiments do not include a control condition where users are aware that the system is actually operated by another human, as it is already well-established in the literature that humans interact with machines differently than they do with other humans. However, as virtual assistants become increasingly intelligent and humans become more accustomed to interacting with them, these differences may begin to shrink, tempering the need for the Wizard of Oz deception. 

\subsubsection{Interactions Between Programmers and Wizards}  
\label{sec:typ}

We can broadly characterize the typical interactions between programmers and wizards in the experiments by considering the four dialogue act dimensions annotated in \Cref{sec:annotation} and \Cref{sec:annotation_results}. We observe that programmers in our study primarily began interactions (i.e. chains of utterances linked by backward-facing dialogue act types) by eliciting information or suggestions from the wizards. These questions most frequently related to five types of API information: patterns (e.g., ``what functions should I use to bring about a specific outcome?''), basic (e.g., ``what is this function's return type?''), functionality (e.g., ``what does this function do?''), structure (e.g., ``what functions are related to this one?''), and examples (e.g., ``show me an example invocation of this function.''). We noted that the majority of programmer utterances that began interactions contained no traceability links to specific API components, while the majority of immediate wizard responses did.

Taken together, while the wizards in our study engaged in a variety of API-related interactions, we found that the majority revolved around identifying the appropriate API components to bring about specific outcomes. We identify five distinct phases in this process: 1) the programmer explaining his/her requirements and/or specifying the type of information requested, 2) the wizard verifying those requirements, 3) the wizard providing one or more information items, 4) the user asking questions about those items, and 5) the wizard answering those questions.

The preponderance of this interaction pattern is due, at least in part, to the specific scenarios implemented in this study; had programmers been performing maintenance tasks, there may have been fewer questions about implementing new patterns and more about understanding existing code. Similarly, had the programmers been granted open access to other documentation resources, they likely would have asked fewer questions geared toward, e.g., basic API information.

\subsubsection{Positive and Negative Programmer Experiences}

We identified a few specific traits in the simulated virtual assistant that programmers appeared to view positively, and a few that they viewed negatively. First, we observed that programmers did not like relying solely on Apiza for API information. By and large, the developers in this study expressed frustration that they were forbidden from accessing typical documentation. While some programmers acknowledged that they were just ``used to'' traditional documentation, they also cited several specific grievances, including the slow response speed, the inability to easily follow links to other components or obtain a high-level view of the API structure, and the fact that responses were sometimes irrelevant, incorrect, or inadequate. 

Second, we observed that programmers remarked positively on the more ``intelligent'' features exhibited by Apiza. In particular, many were impressed by Apiza's ability to provide context-specific recommendations and synthesize information when the documentation was incomplete.  While the programmers were frustrated with \textit{how} Apiza communicated information, they were impressed by its ability to determine \textit{what} to communicate. One programmer described how ``[it] was very cool how perfectly Apiza understood what I wanted to know and even offered code samples.'' Another programmer expressed, ``I would say that an AI would come in handy when I don't actually know what I want or what to do. Or rather, the only times that I could see myself using it would be times that I usually try to ask a person.''

Third, many programmers in our study shared a key complaint about the simulated virtual assistant: it was too slow. As one programmer summarized, ``the speed was by far the most frustrating part.'' A real VA has an advantage over a human wizard, in that it can rapidly search through large amounts of data; however, features that involve accessing online resources or running neural models may cause severe delays. Virtual assistant developers will have to balance the desire to include sophisticated features against the need for the system to generate a speedy response.

\subsubsection{Additional Considerations}

We present some final considerations for researchers and virtual assistant developers as they pursue this research direction. First, they should consider that many IDEs incorporate advanced features, such as autocompleting function names and displaying information regarding parameters and return types. In cases where those features are already available within an IDE, they may be of lesser value in a virtual assistant; this may affect decisions regarding which dialogue actions should be made available to the system. At the same time, if developers are targeting specialized applications or user types for which accessibility to other IDE features may be limited, it may be sensible to include such features in a virtual assistant even if they are redundant.

Second, researchers and virtual assistant developers should consider different ways in which online resources may be incorporated by a virtual assistant. In our studies, wizards occasionally directed programmers to websites related to the programmers' queries. We also frequently observed wizards struggle to determine how to help the programmers implement certain design patterns; in many of those cases, the wizards could have easily found the answer by searching the internet. A virtual assistant can take advantage of the huge amount of API information available on question-answering websites like StackOverflow, in projects hosted on sites like GitHub, and in tutorials around the web.

Third, researchers and virtual assistant developers will have to make a number of important decisions when implementing a user interface (UI). In particular, the input and output modalities are likely to have a strong impact on the overall usability of the system. The programmers and wizards in our experiments communicated solely through a text-based interface, but there are several alternative forms of input (such as voice and mouse/touchscreen input) and output (such as computer speech and automatic navigation to relevant documentation/resources) that we did not explore. The text-based system was ideal for this study, as it did not restrict the types of questions or information that either party could share. However, it did limit the quantity of information a wizard could reasonably share in a single turn, and many users found it inconvenient compared to standard documentation navigation. Future researchers and developers should consider a broad range of suitable UIs, and their effects on usability.

\subsection{Recommendations}
\label{sec:recs}

Following the framework described in \Cref{sec:strats}, developers may use the data from this study to create dialogue strategies for a virtual assistant for API usage via reinforcement learning. Here, we discuss the next steps in this process and give specific recommendations regarding the application of the Wizard of Oz data.

\smallskip
\vspace{.1cm}

\begin{enumerate}[leftmargin=0cm,itemindent=.5cm,labelwidth=\itemindent,labelsep=0cm,align=left, listparindent=1.5em]
\item \textit{Considering Core Functionality.} 

Virtual assistant developers must first decide on the core functionality they would like their systems to exhibit. The wizards in our study performed a number of distinct activities: they helped users identify relevant API components, they answered questions about those components, they fetched code examples from the source code and online resources, and more. Choosing the particular task(s) a virtual assistant should be able to complete will guide the rest of its development.

To make this determination, developers should consider the programmer behaviors and interaction patterns observed in this study, as well as research from the broader software engineering field. Below, we give examples geared toward the development of dialogue strategies specifically designed for the task of identifying API components that fulfill users' information needs (similar to the frequent interaction pattern described in \ref{sec:typ}).

\smallskip
\vspace{.1cm}

\item \textit{Identifying system and user actions.} 

After determining the core functionality of the target system, developers must establish the sets of dialogue actions to be made available to the system and the user. These action sets are directly informed by the desired system functionality, and must reflect decisions that the system will need to make to enable sufficiently complex dialogue.

As mentioned in \Cref{sec:woz}, the process of determining appropriate action sets can be expedited by utilizing the data presented in this study. In short, this process involves first identifying all utterances that may be related to the target functionality. In the hypothetical system to recommend API components, developers might begin by looking at user utterances of the ELICIT-OFFER-OR-SUGGESTION illocutionary type with the PATTERNS label in the API dimension (as these utterances are largely associated with the target functionality), and then collecting all utterances that connect to these via backward-facing dialogue act types.

Next, the developers would look at the combinations of dialogue act types that occur across all four dimensions in this subset of utterances, and attempt to group together all combinations that relate to target functionality. For example, while some user utterances of the ELICIT-OFFER-OR-SUGGESTION type do not contain traceability labels, others reference specific API components that the user believes are related to the desired functionality. The developers may choose to account for this observed behavior by implementing two specific search methods: one that uses a simple keyword query, and one that considers direct structural links between API components in addition to a keyword query. Therefore, the developers could assign a new ``BASIC-QUERY'' label to the first dialogue act combination, and a ``RELATED-QUERY'' to the second (as well as matching ``BASIC-SEARCH'' and ``RELATED-SEARCH'' labels for corresponding system actions).

At this point, it may become apparent that certain observed or desired functionality cannot be derived from the existing annotations with a sufficient level of granularity. For instance, in a virtual assistant that suggests relevant functions, it may be important to differentiate between actions in which the user elicits \textit{all relevant functions} and those in which the user elicits \textit{the most relevant function}. In the current annotation scheme, both of these would simply be labeled with the ELICIT-OFFER-OR-SUGGESTION illocutionary dialogue act type, but based on their observations, the developers may decide to use separate ``QUERY-ALL'' and ``QUERY-BEST'' labels.

Ultimately, the developers should be able to assign actions from the system and user action sets to all relevant utterances in the corpus, as the annotated corpus will later be used to build a simulated learning environment.

\smallskip

\item \textit{Narrowing down the state space.}

The ``dialogue state'' refers to a reinforcement learning agent's beliefs and view of its environment, and the ``state space'' is the set of all possible states. The state space is defined by a set of features, and each individual state corresponds to a particular set of values for those features. Choosing an appropriate, narrow set of features for the state space facilitates the learning process by enabling informed decision-making. 

Virtual assistant developers can use an annotated corpus to identify which task- and dialogue-specific features actually influenced the wizards' dialogue strategies, in order to determine which ones to include in the state space. To do so, they must first generate a set of candidate state space features, and then prune down the most useful ones. Reiser and Lemon~\cite{rieser2011reinforcement} demonstrate how, for candidate features that are readily observable in the corpus, this can be done by determining which features are most predictive of wizard's dialogue acts.

Candidate features to be included in the state space for the target API virtual assistant include dialogue length, how many questions the user has asked, whether the user has repeated the current question, the previous user and system actions, whether a specific API component has been identified, the number of components in the API that appear to be relevant to a user's query and how relevant those components appear to be, the number of web resources that appear to be relevant to a user's query and how relevant those appear to be, a belief about the user's ongoing programming activity (bugfixing, refactoring, implementing a new method, etc.), whether the user's request has been satisfied, and many more.
\smallskip

\item \textit{Building a Simulated Learning Environment.}

Once the action sets and state space are established, the next step is to create a model that represents the simulated environment and dictates how different actions lead to different states. In the context of reinforcement learning for dialogue strategy, the primary component of this model is referred to as a ``user simulator.'' The user simulator emulates how a user would react to the system's action at a given point in a dialogue. In many ways, the user simulator can be just as complex as the agent it is used to train; it has its own set of actions, goals, and constraints that can change over the course of a dialogue. It is common for the user simulator to use manually crafted heuristics to update its goals and constraints and to use supervised machine learning to choose an appropriate action. More rules are then needed to define how the agent's state changes in response to the simulated user's action.

Rieser and Lemon~\cite{rieser2011reinforcement} demonstrate a fairly straightforward user simulator that keeps track of whether the agent has completed certain tasks and uses a simple bigram model (trained on labeled Wizard of Oz data) to generate user actions. Once again, a virtual assistant developer should consider the interactions that took place in the real Wizard of Oz dialogues and determine how sophisticated a user simulator needs to be for a particular application. For instance, a simple user simulator designed to train a virtual assistant to identify relevant API components may choose a target function and then ask questions and provide information related to that function. But the developer may also want the virtual assistant to account for scenarios in which the user asks for a component that does not exist, or scenarios in which the user initially wants a certain component, but is satisfied by a similar component. Any desired functionality in the virtual assistant must be reflected in the user simulator.

\smallskip

\item \textit{Defining the reward function.}

The reward function enables the reinforcement learning agent to learn optimal strategies by assigning point values to specific actions and outcomes to encourage and discourage certain behaviors. For example, agents are usually penalized a small amount for each turn in a task-oriented dialogue to encourage them to complete the task as quickly as possible. They are also generally rewarded or penalized depending upon whether they complete their task within a specified number of time steps. However, in the context of task-oriented dialogues, there are other behaviors that a virtual assistant developer may wish to discourage; for instance, users may not like it if the virtual assistant shares too much information in a single dialogue turn, even though doing so might increase the likelihood of the agent completing its task. Developers must decide which dialogue features to include in the reward function, and how to balance them.

While this task can be somewhat reliant on trial and error, Reiser and Lemon~\cite{rieser2011reinforcement} suggest using Wizard of Oz studies to help define a reward function by looking at the correlation between dialogue features and the experimental performance metrics. For instance, they demonstrate how linear regression can sometimes be used to estimate the effect of dialogue length on user satisfaction. It can also be a useful diagnostic tool to determine which features not to include in the reward function (i.e. those with little correlation to the metrics).

In the context of the present study, future developers may want to consider how well features like interaction length, specific dialogue act types, and progress towards goal completion correlate with the observed task completion rates (\Cref{sec:task_completion}) and user satisfaction ratings (\Cref{sec:user_satisfaction}), but they should remain mindful that these correlations may be confounded by other, uncontrolled factors in the experiments. 

\end{enumerate}

\section{Conclusion}
\label{sec:conclusion}

Virtual assistants for programmers have not been widely researched, despite recent advancements in virtual assistant technology and calls for more intelligent tools in software engineering. This is largely due to the lack of publicly-available datasets that can be used to understand which programming tasks would be high-value targets for virtual assistants and to train task-specific dialogue systems.

In this paper we laid the groundwork for a virtual assistant for API usage. First, we presented the methodology and results of Wizard of Oz experiments designed to simulate interactions between a programmer and a virtual assistant for API usage. Then, we annotated the dialogue acts in the programmer-wizard interactions along four dimensions: illocutionary dialogue act type, API dialogue act type, backward-facing dialogue act type, and traceability. Finally, we discussed the implications of our study on future virtual assistant development.

\smallskip
We have made all data related to the experimental design, experimental results, and dialogue act annotations available via an online Appendix:

\smallskip

\noindent
\url{https://github.com/ApizaCorpus/ApizaCorpus}
\section*{Appendix A}
\label{app:participants}
Tables 5 and 6 summarize all responses to the entry surveys completed by participants in the API Wizard of Oz study. All entry surveys and responses are given in the online Appendix.

\begin{savenotes}
    \begin{table*}[t]
    \caption{Summary of the entry surveys for programmers in the API Wizard of Oz study. Two programmers in the libssh scenario and one in the Allegro scenario did not submit responses to the survey, and one in the Allegro scenario did not report familiarity with the API domain. Note that the survey questions relating to regular C use and native language were only asked of programmers in the Allegro scenario. Additionally, programmers in the Allegro scenario were asked to provide an exact number of years for questions regarding programming experience, while those in the libssh study were asked to select from the ranges in the table.}
    \centering
    \begin{tabular}{llcccc}
    \toprule
        \multirow{2}{*}{Survey Topic} & \multirow{2}{*}{Survey Response} & \multicolumn{2}{c}{\# Programmers} \\ 
        &  & libssh Scenario & Allegro Scenario \\
    \midrule
        \multirow{3}{*}{Programming Experience (Any Language)} & \textless 2 Years & 1 &0 \\
        & 2-10 Years & 7& 5\\
        & \textgreater 10 Years &5 &9 \\
    \midrule
        \multirow{3}{*}{Programming Experience (C Language)} & \textless 2 Years & 7 &2 \\
        & 2-10 Years & 5& 6 \\
        & \textgreater 10 Years &1 & 6 \\
    \midrule
        \multirow{3}{*}{Familiarity with API Domain} & Novice & 6 & 11 \\
        & Intermediate & 7& 1 \\
        & Expert & 0& 1  \\
    \midrule
        \multirow{2}{*}{Uses C Regularly} & Yes & N/A & 9 \\
        & No & N/A & 5 \\
    \midrule
        \multirow{2}{*}{Native Language} & English & N/A & 8 \\
        & Other & N/A & 6 \\
    \bottomrule
    \end{tabular}
    \label{tab:progexp}
    \end{table*}
\end{savenotes}{}

\begin{savenotes}
    \begin{table*}[t]
    \caption{Summary of the wizards' programming experience and scenario participation in the API Wizard of Oz study.}
        \centering
        \begin{tabular}{ccccc}
        \toprule
            \multirow{2}{*}{Wizard ID} & \multicolumn{2}{c}{Programming Experience} & \multicolumn{2}{c}{\# Sessions}  \\
            &Any Language&C Language &libssh Scenario&Allegro Scenario\\
        \midrule
            1 & 2-10 Years & 2-10 Years & 6 & 4\\
            2 & 2-10 Years & 2-10 Years & 3 & 2\\
            3 & 2-10 Years & \textless 2 Years& 2 & 3\\
            4 & 2-10 Years & \textless 2 Years& 3 & 3\\
            5 & 2-10 Years & \textless 2 Years& 1 & 0\\
            6 & 2-10 Years & 2-10 Years& 0 & 3\\
        \bottomrule
        \end{tabular}
        \label{tab:wizexp}
    \end{table*}
\end{savenotes}{}

\section*{Appendix B}
\label{app:labels}

All illocutionary, API, and backward-facing dialogue act types are listed in Tables \ref{tab:amilabels}, \ref{tab:apilabels}, and \ref{tab:bflabels}, respectively. A description is given for each dialogue act type, as is an example utterance from the Wizard-of-Oz API usage corpus. The complete sets of API component labels for the traceability dimension are available in the online Appendix in the \texttt{traceability\_labels/} folder.

\begin{table*}[t!]
    \centering
    \caption{Description of 12 Illocutionary dialogue act types derived from the AMI Meeting Corpus~\cite{mccowan2005ami}. We have excluded three dialogue act types from the original annotation scheme (``STALL'', ``FRAGMENT'', and ``BACKCHANNEL'') that were not relevant for written communication.} 
    \begin{tabularx}{\textwidth}{p{4cm}XX}
        \toprule
        Dialogue Act Type & Description & Example Utterance\\
        \midrule
        INFORM & A statement intended to convey information to the other speaker.  & \textbf{Programmer}: al\_get\_keyboard is failing with ``Assertion `new\_keyboard\_driver' failed''.\\[12pt]

        ELICIT-INFORM & A request for the other speaker to share information.  & \textbf{Programmer}: Can I pass ssh\_connect a hostname?\\[7pt]

        ASSESS & An evaluation of something that is being discussed. & \textbf{Wizard}:  Great!\\[7pt]

        ELICIT-ASSESSMENT & A request for the other speaker to evaluate something that has been said or done so far. & \textbf{Wizard}:  Does this look useful?\\[12pt]

        OFFER & An act in which the speaker expresses an intention relating to possible actions that he or she may take. & \textbf{Programmer}:  Well I'll do an experiment and let you know \\[12pt]

        BE-POSITIVE & Any act that is intended to express positive feelings toward the other speaker. & \textbf{Programmer}: Thanks for your help today. \\[12pt]

        BE-NEGATIVE & Any act that is intended to express negative feelings toward the other speaker. & \textbf{[N/A]}\\[12pt]

        SUGGEST & An act in which the speaker expresses an intention relating to possible actions that the other speaker may take. & \textbf{Wizard}:  To install a keyboard driver, use al\_install\_keyboard.\\[17pt]

        ELICIT-OFFER-OR-SUGGESTION & An act in which the speaker expresses a desire for the other speaker to make an offer or suggestion & \textbf{Programmer}: How do I register key events?\\[12pt]

        COMMENT-ABOUT-UNDERSTANDING & A comment indicating whether or not the speaker understood a previous utterance. & \textbf{Wizard}: I don't understand your question\\[12pt]

        ELICIT-COMMENT-ABOUT-UNDERSTANDING & A request for the other speaker to indicate whether or not he or she understood a previous utterance. &\textbf{Programmer}: Apiza are you there \\[12pt]

        OTHER & A ``bucket'' class for other speaker intentions, such as stalls, text fragments, and unclear intentions. &\textbf{Wizard}: one second, parsing source code...\\
        \bottomrule
    \end{tabularx}
    \label{tab:amilabels}
\end{table*}

\begin{table*}[t!]
    \centering
    \caption{Description of 12 API dialogue act types derived from the API knowledge taxonomy~\cite{maalej2013patterns}. We have added a ``BASIC'' label in place of a ``NON-INFORMATION'' label.} 
    \begin{tabularx}{\textwidth}{lXX}
        \toprule
        Dialogue Act Type & Description & Example Utterance \\
        \midrule

        FUNCTIONALITY & \emph{What} the API does in terms of functionality, including description of parameters, return values, and exceptions. & \textbf{Programmer}: why would al\_load\_sample crash?\\[12pt]

        CONCEPTS & The \emph{meaning} of terms used to name or describe an API element, or \emph{design or domain concepts} used or implemented by the API. & \textbf{Programmer}: what is the x y orientation of the display \\[20pt]

        DIRECTIVES & Information relating to what users are \emph{allowed/not allowed} to do with the API element. Directives are clear contracts. & \textbf{Wizard}: al\_install\_audio must have been called first.\\[12pt]

        PURPOSE & The \emph{purpose} of providing an element or the \emph{rationale} of a certain design decision. & \textbf{Wizard}: An ALLEGRO\_COLOR structure describes a color in a device independent way. \\[12pt]

        QUALITY & Quality attributes of the API, also known as non-functional requirements, including the \emph{performance} implications and information about the API’s \emph{internal implementation} that is indirectly related to its observable behavior. & \textbf{Wizard}: Playback may fail because all the reserved sample instances are currently used.\\[30pt]

        CONTROL & How the API (or the framework) manages the flow of control. For example, what events cause a certain callback to be triggered, or the order in which API methods will be automatically called by the framework itself. & \textbf{Programmer}: when a user presses a key, is a corresponding event placed in the event queue? \\[30pt]

        STRUCTURE & The internal organization of a compound element (e.g. important classes, fields, or methods), information about type hierarchies, or how elements are related to each other. & \textbf{Wizard}: Members include acceptForward(int timeout\_ms), connect(), disconnect(), getAuthList()...\\[20pt]

        PATTERNS & How to accomplish specific outcomes with the API, for example, how to implement a certain scenario, how the behavior of an element can be customized, etc. & \textbf{Wizard}: To send an end of file on the channel, use ssh\_channel\_send\_eof\\[20pt]

        EXAMPLES & Code examples demonstrating how to use and combine elements to implement certain
functionality or design outcomes & \textbf{Programmer}: can you give me an example of ssh\_channel\_read \\[20pt]

        ENVIRONMENT & Aspects related to the environment in which the API is used, but not the API
directly, e.g., compatibility issues, differences between versions, or licensing information. & \textbf{Programmer}: When was libssh Version 1.x released? \\[20pt]

        REFERENCES & Any pointer to external documents or mentions of other documents (such as standards or manuals). Note that this tag does not account for internal API references (i.e. references to other API components).& \textbf{Programmer}: website of libssh documentary\\[30pt]

        BASIC & ``Boilerplate'' information such as component names, return types, and parameters, as well as broad requests for information relating to certain API components. & \textbf{Programmer}: Can I have more information about the ALLEGRO\_KEYBOARD\_EVENT object?\\
        \bottomrule
    \end{tabularx}
    \label{tab:apilabels}
\end{table*}

\begin{table*}[t!]
    \centering
    \caption{Description of 7 backward-facing dialogue act types derived from the AMI Meeting Corpus~\cite{mccowan2005ami}. We have added three labels to the original set to account for frequent inter-utterance relationships that occurred in the context of written API dialogues: CONTINUE, REPEAT, and FOLLOW-UP.} 
    \begin{tabularx}{\textwidth}{lXX}
        \toprule
        Dialogue Act Type & Description & Example Utterance  \\
        \midrule
        POSITIVE & An act that supports the intention of a prior utterance, for instance, by reacting positively to it, accepting or agreeing with it, indicating it has been understood, or providing what the source is attempting to elicit.  & \textbf{Wizard}: To connect in libssh, you can use the function ``int ssh\_connect(ssh\_session session)''\\[30pt]
        NEGATIVE & An act rejecting a prior utterance, for instance, by presenting an objection to it, countering the source with an alternative the speaker prefers, or refusing to provide what the source is attempting to elicit. & \textbf{Programmer} (in response to a suggestion): I tried that\\[30pt]
        PARTIAL & An act the partially supports a prior utterance, but rejects it in some aspects, for instance by agreeing with part of a suggestion or providing part of what the source is attempting to elicit. & \textbf{Programmer}: tell the easiest one \\[30pt]
        UNCLEAR & An act that expresses genuine uncertainty about a prior utterance, for instance, by saying that speaker is unsure whether or not a suggestion is a good idea or whether some information is true, or by expressing an inability to provide what the source is attempting to elicit & \textbf{Wizard}: I am unsure\\[38pt]
        REPEAT & An act that repeats or rephrases a prior utterance by the same speaker in order to, for instance, emphasize a point, redirect the dialogue, ensure that the prior utterance was successfully received, or repair a typo or other mistake. & \textbf{Programmer} (after a previous utterance ``How do I set the hostname for an SSH session?''): How do I set a hostname on an SSH session?  \\[30pt]
        FOLLOW-UP & An act that neither supports nor rejects a prior utterance, but arises from information conveyed in that utterance, for instance, by posing a follow-up question, or suggesting a related topic.  & \textbf{Programmer} Just to confirm, is the ssh\_session type a pointer type? \\[30pt]
        CONTINUE & An act that serves as a direct continuation of a prior utterance, but has either been delivered in a separate message or serves a role distinct from the prior utterance (for instance, by elaborating on a previously-introduced topic). & \textbf{Wizard} (immediately after suggesting the function al\_load\_bitmap): ALLEGRO\_BITMAP *al\_load\_bitmap(const char *filename)\\
        \bottomrule
    \end{tabularx}
    \label{tab:bflabels}
\end{table*}


\section*{Acknowledgments}

The authors would like to sincerely thank the participants in the Wizard of Oz experiments, as well as the anonymous reviewers whose reccomendations have greatly improved the manuscript.

\bibliographystyle{IEEEtran}
\bibliography{IEEEabrv,biblio}

\vskip 0pt plus -1fil

\begin{IEEEbiography}
 [{\includegraphics[width=1in,height=1.25in,clip,keepaspectratio]{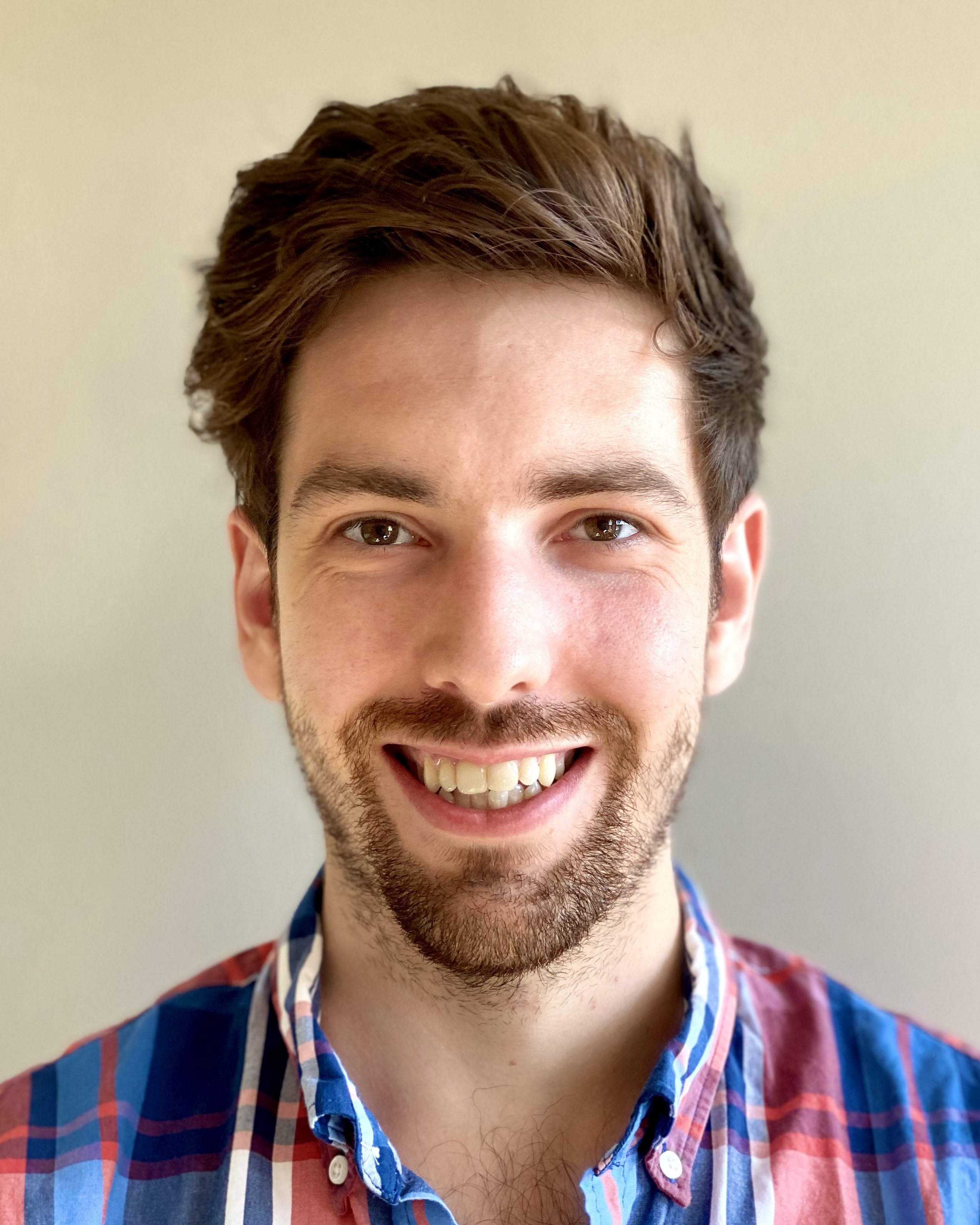}}]{Zachary Eberhart} is working toward the PhD degree at the University of Notre Dame advised by Dr. Collin McMillan.  His research is in software engineering with a focus on human-computer interaction, code reuse, and program comprehension.
\end{IEEEbiography}

\vskip 0pt plus -1fil

\begin{IEEEbiography}
 [{\includegraphics[width=1in,height=1.25in,clip,keepaspectratio]{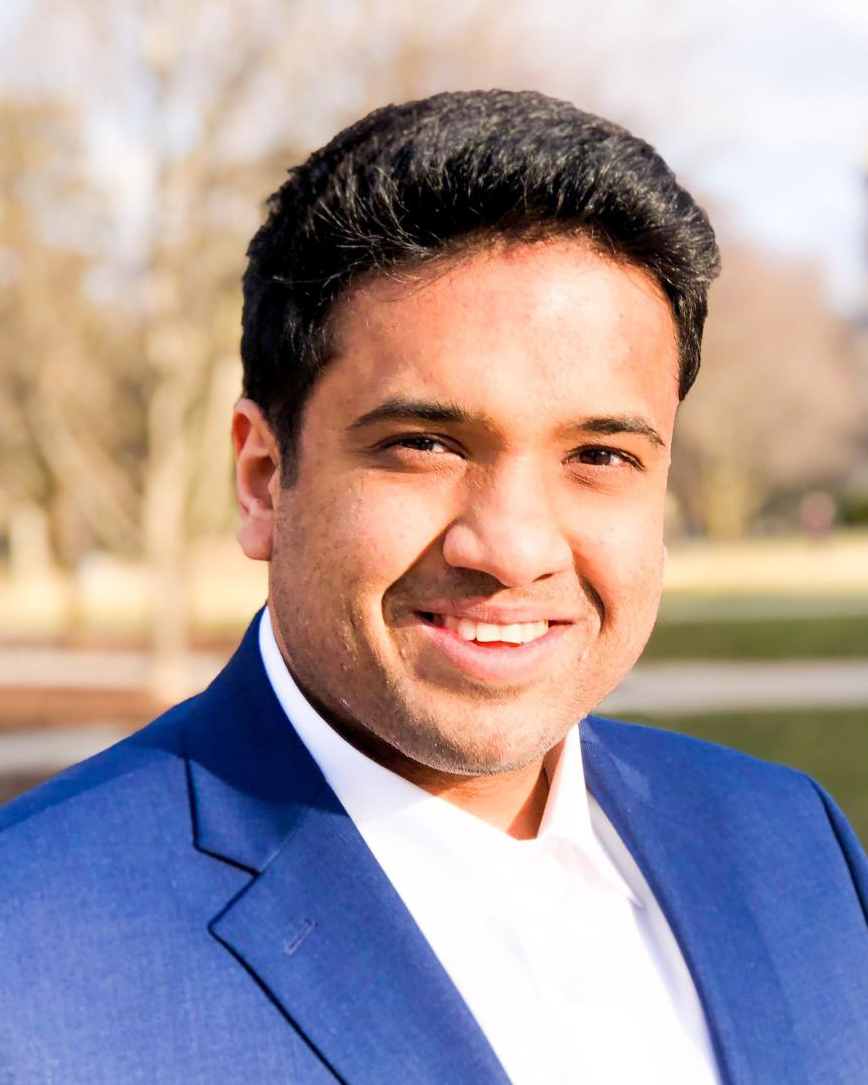}}]{Aakash Bansal} is working toward the PhD degree at the University of Notre Dame advised by Dr. Collin McMillan.  His research is in software engineering with a focus on source code summarization and program comprehension.
\end{IEEEbiography}
\vskip 0pt plus -1fil
 
\begin{IEEEbiography}
 [{\includegraphics[width=1in,height=1.25in,clip,keepaspectratio]{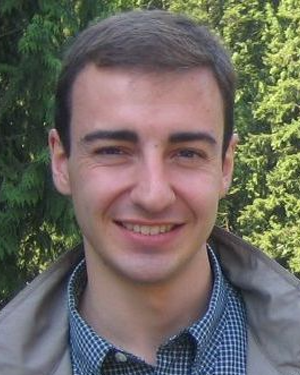}}]{Collin McMillan} received the PhD degree from the College of William \& Mary, in 2012, focusing on source code search and traceability technologies for program reuse and comprehension. He is an associate professor with the University of Notre Dame. Since joining Notre Dame, his work has focused on source code summarization. His work has been recognized with multiple best paper and distinguished paper awards, and the NSF CAREER award. He is a member of the IEEE.
\end{IEEEbiography}

\end{document}